\documentclass[12pt,preprintnumbers,nofootinbib]{revtex4}

\usepackage{amsmath,amsthm,amssymb,epsf}
\usepackage{mathbbol,bbm}
\usepackage{graphicx}
\usepackage{wasysym}
\usepackage{subfigure}

\newcommand{\beqa}{\begin{eqnarray}}
\newcommand{\eeqa}{\end{eqnarray}}
\newcommand{\f}{\begin{equation}}
\newcommand{\ff}{\end{equation}}

\newcommand{\bean}{\begin{eqnarray*}}
\newcommand{\eean}{\end{eqnarray*}}
\newcommand{\ra}{\rightarrow}

\newcommand{\pa}{\partial}
\newcommand{\mc}[1]{{\mathcal{{#1}}}}

\newcommand{\cln}{\colon}

 \def\HH{{\cal H}}

\def\be{\begin{equation}} \def\ee{\end{equation}}

\begin{document}

\title{Statistical Mechanics of Graphity Models}
\author{Tomasz Konopka}
\affiliation{ITP, Utrecht University, Utrecht 3584 CE, the
Netherlands}

\preprint{ITP-UU-08/28}\preprint{SPIN-08/21}

\begin{abstract}
Graphity models are characterized by configuration spaces in which
states correspond to graphs and Hamiltonians that depend on local
properties of graphs such as the degrees of vertices and numbers
of short cycles. As statistical systems, graphity models can be
studied analytically by estimating their partition functions or
numerically by Monte Carlo simulations. Results presented here are
based on both of these approaches and give new information about
the high- and low-temperature behavior of the models and the
transitions between them. In particular, it is shown that matter
degrees of freedom must play an important role in order for the
low-temperature regime to be described by graphs resembling
interesting extended geometries.
\end{abstract}
\maketitle

\section{Introduction}

The idea of describing space (or spacetime) as a product of a
dynamical process involving some discretized elements resonates
with many approaches to studying the unification of quantum
mechanics with gravity
\cite{Henson:2006kf,Oriti:2007qd,Baez:1999sr, Jourjine:1984jz,
Rideout:1999ub, Requardt:1999db,QCHistories,
Barcelo:2005fc,Volovik:2007vs,Dreyer:2006pp,
DynTrianLorentzQG,Ambjorn:2005qt,Ambjorn:2004qm,
LQG1,LQG2,Giesel:2006uj}. ``Graphity'' refers to a class of models
in which the degrees of freedom represent dynamical graphs
\cite{graphity1,graphity2} and whose Hamiltonians depend only on
minimal information encoded in graphs such as the degree of the
vertices and cycles formed by the edges. It has been argued that
such models can be useful in studying how an extended and/or
latticelike geometry might emerge from a background independent
dynamical system.

Several approaches to studying quantum geometries are based on
general or directed graphs and graphity models share some features
with them while differing in others. In contrast to some setups
defined primarily by a set of dynamical rules
\cite{Henson:2006kf,Rideout:1999ub,Requardt:1999db,QCHistories},
graphity models have Hamiltonians, which allow them to be
interpreted as statistical mechanical systems. In contrast with
approaches starting from a quantization ansatz for general
relativity \cite{DynTrianLorentzQG,Ambjorn:2005qt,Ambjorn:2004qm,
LQG1,LQG2,Giesel:2006uj}, graphity models are only based on
graph-theoretic notions and are not specific to a particular
dimension or embedding manifold. They are in nature most similar
to dynamical triangulations
\cite{DynTrianLorentzQG,Ambjorn:2005qt,Ambjorn:2004qm} in the
sense that their macroscopic properties are determined by the
collective interactions between a large number of small elements.
The graph models are also amenable to simulation just like
dynamical triangulations
\cite{DynTrianLorentzQG,Ambjorn:2005qt,Ambjorn:2004qm}. However,
since not all general graphs correspond to triangulations,
graphity models have a larger configuration space and their
defining Hamiltonian cannot be easily mapped to a discretized
version of the action of general relativity.

The motivation for studying such general systems comes from asking
to what extent geometric notions such as the continuum, dimension,
or macroscopic locality, should be considered as fundamental (as
they are in all approaches that postulate the existence of a space
or space-time manifold) and to what extent they may be only
approximate or emergent. One way of probing these issues is to
consider a setup in which none of these notions is obviously true.
Graphity models are an attempt in this direction that assume only
the existence of a Hamiltonian function $H$ determined by
micro-local conditions of graphs (for a discussion of the
distinction between macro and micro locality, see
\cite{Markopoulou:2006qh}.)

It has been argued that graphity models can behave very
differently at high and low energies \cite{graphity1,graphity2}.
At low energies, the preferred graph state can represent an
emergent geometry, while at high energies, the system is expected
to be disordered and highly connected. The purpose of this paper
is to study the statistical mechanics of the models in order to
learn more about the nature of the low-temperature regime. The
main finding is that while the most simple models do generate
extended graphs at low temperatures, these graphs are effectively
only one dimensional. However, when matter degrees of freedom that
are sensitive to graph boundary conditions are added, the
low-energy behavior can dramatically change and various homogenous
latticelike graphs may appear. It thus appears that matter, as has
been noticed in other contexts as well
\cite{Dreyer:2006pp,KalyanaRama:2006xg,Bilke1,Bilke2}, should be
regarded as an important, if not crucial, factor in discussions of
emergent geometries.

The definition of graphity models is reviewed in the next section.
A distinction is made between a basic, or bare-bones, model of
pure graphs and more sophisticated versions that can incorporate
matter degrees of freedom in the form of gauge fields as well. The
interpretation of the models as statistical systems is also
introduced laying the ground for much of the following discussion.
Section \ref{s_statmech1} deals with the partition function of the
basic model. This discussion is detailed and involves identifying
the types of graphs that can contribute significantly to the
system's partition function. The discussion should be seen as
complementary to the results of Monte Carlo simulations presented
in Sec. \ref{s_simulations}. Indeed, it is the simulations that
motivate and at the same time justify the study of particular
types and the omission of other types of graphs in the theoretical
section. As such, the analysis in Sec. \ref{s_statmech1} is an
attempt to provide analytical understanding of the results of the
numerical simulations. And in turn the analytical work suggests
what kind of numerical studies can be of potential interest.

The results of the Monte Carlo simulations in Sec.
\ref{s_simulations} indicate that the basic graphity model is
insufficient to generate extended geometries with dimension
greater than one in the low-energy regime. The emerging graphs are
either chainlike or treelike. It is shown in this section,
however, that supplementing the basic model with a condition for
graph homogeneity does lead to very different and interesting
results. This observation motivates the study of graph models with
matter in Sec. \ref{s_Zparticles}. The discussion there mirrors
that in Sec. \ref{s_statmech1} and demonstrates that under certain
quite general assumptions regarding the energy spectrum of the
matter content, a model with a realistic matter component may
indeed produce extended geometries at low temperature. A summary
of all the findings is presented in Sec. \ref{s_discussion}.


\section{Graphity Models \label{s_models}}

Graphity \cite{graphity1,graphity2} is a name for a class of
models in which states are associated with graphs and the
configuration space is the space of all possible simple graphs
with a given number of nodes $N$. The models are defined in a
quantum mechanical context so that the configuration space is
actually a Hilbert space. Indeed the Hamiltonian of the model is
most naturally formulated as a quantum mechanical operator rather
than as a classical object. For this reason, the models can also
be referred to as ``quantum'' graphity.

The total Hilbert space for a graphity model can be decomposed as
\be \label{HHtot} \HH_{total} = \bigotimes^{N(N-1)/2} \!\!\!\!
\HH_{edge} \; \bigotimes^{N} \HH_{vertex}. \ee The tensor products
of $\HH_{vertex}$ and $\HH_{edge}$ corresponds to putting degrees
of freedom on each vertex and every possible edge connecting the
vertices. Specific graphity models are defined by particular
choices for the $\HH_{vertex}$ and $\HH_{edge}$, along with an
appropriate Hamiltonian acting on the resulting $\HH_{total}$.

\subsection{Basic Model}

In the basic model described in \cite{graphity2}, the Hilbert
space associated with the vertices is trivial and the Hilbert
space associated with each edge is spanned by two vectors \be
\label{HHedge} \HH_{edge} = \mathrm{span}\{ \, |\,0\,\rangle, \,
|\,1\,\rangle \}. \ee The total Hilbert space is therefore \be
\label{HHtotalsmall} \HH_{total} = \bigotimes^{N(N-1)/2} \!\!\!\!
\HH_{edge}. \ee Denoting states in each copy of $\HH_{edge}$ as
$|n_{ab}\rangle$ with $n_{ab}=0,1$, the total Hilbert space can be
said to be spanned by the product states \be \label{Htotbasis}
\HH_{total} = \mathrm{span}\{ |n_{12}\rangle \otimes
|n_{13}\rangle \otimes |n_{23}\rangle \otimes \cdots \} \ee A
general state in this space is a superposition of the basis states
with complex coefficients.

Interpreting the states $|\,1\,\rangle$ and
$|\,0\,\rangle$ as conveying whether or not, respectively, a link
is present between the two vertices, each of the basis states (\ref{Htotbasis}) can be associated with a graph configuration
or a graph diagram. Thus, the total Hilbert space
(\ref{HHtotalsmall}) can be decomposed as \be \HH_{edges} =
\bigoplus_G \HH_{G} \ee where $\HH_{G}$ is a space spanned by a
single basis state and corresponds to a single
graph $G$. Because the vertices are labeled and are thus
distinguishable, the tensor summation over $\HH_G$ can contain
multiple graphs $G$ that are isomorphic to each other.

As argued in \cite{graphity2}, a Hermitian Hamiltonian $H$ acting
on $\HH_{total}$ can be used to associate an energy $E(G)$ with a
graph states $|\psi_G\rangle$ through the relation \be
\label{EGexpectation} E(G) = \langle \psi_G | \cln H\cln
|\psi_G\rangle. \ee The form of the Hamiltonian is restricted by
requirements of general graph locality and can thus depend only on
a few properties of the graph. The basic model described in
\cite{graphity2} uses a Hamiltonian of the form \be \label{HVB} H
= H_V + H_B +H_{int}\ee where $H_V$ depends on the degree of
vertices, $H_B$ depends on the cycle structure of a graph, and
$H_{int}$ is an interaction term. Each of the terms in the
Hamiltonian can be implemented using creation and annihilation
operators acting on $\HH_{total}$ \cite{graphity2}. For the
purposes of this paper, however, it is sufficient to describe only
the effective properties of the energies $E$ associated with these
terms.

The valence term $H_V$ assigns an energy to the graph according to
\be E_V = g_V \sum_a e^{p\, (v(a) - v_0)^2}. \ee Here $g_V$ is a
positive coupling constant, $p$ is some positive real number,
$v(a)$ is the degree of vertex $a$, and $v_0$ is positive integer
that determines the preferred valence of each vertex in the graph.

The other term $H_B$ is such that its contribution to the energy
can be written as \be E_B = \sum_a E_B(a)\ee with \be \label{EBa}
E_B(a) = - \sum_{L=3}^{N(N-1)/2} g_B(L) P(a,L). \ee The last
expression is a contribution that corresponds to a single vertex
$a$ and that depends on the number of cycles $P(a,L)$ of length
$L$ that pass through that vertex. A cycle is defined as a closed
walk along the edges of a graph in which no edge is traversed more
than once (see \cite{graphity2} for more details.) The lower and
upper limits on the sum are a consequence of this particular
definition of a cycle.

The effective coupling $g_B(L)$ associated with cycles of each
length is \be \label{gbLdef} g_B(L) = g_B\frac{r^L}{L!} \ee with
$g_B$ and $r$ some constants. The form of this effective coupling
plays a very important role in the properties of graphity models.
The interplay between the exponential in the numerator and the
factorial in the denominator of (\ref{gbLdef}) means that the
effective coupling increases with $L$ for small $L$ but decreases
very rapidly for large $L$. The rapid falloff implies that the
energy $E_B$ is primarily determined by the number of short cycles
in a graph. The total energy (\ref{EBa}) may be well approximated
by restricting the summation over $L$ to the range $0<L<L_{max}$
as follows: \be E_B(a) \simeq - \sum_{L=3}^{L_{max}} g_B(L)
P(a,L). \ee The maximal cycle length $L_{max}$ must be determined
for each model depending on the given parameters $v_0$ and $r$ and
on what accuracy the energy is to be evaluated.

The turnover between increasing and decreasing $g_B(L)$ occurs at
some length $L_*$ determined by the parameter $r$. When the
effective coupling $g_B(L)$ is multiplied by the number of cycles
$P(a,L)$, which usually increases with $L$, the maximum value of
the product occurs at another length $L_{**}$ determined by both
$r$ and the type of graph used to extract $P(a,L)$. Both scales
$L_*$ and $L_{**}$ are smaller than $L_{max}$. It is important
that none of these cycle length scales depend on the number of
vertices in the graph $N$. Therefore, since only short cycles are
important, the energy function can be said to be
quasi-(micro)local.

Other important properties of (\ref{gbLdef}) are determined by the
signs of $r$ and $g_B$. For $r>0$ and $g_B>0$, the effective
coupling is positive definite; when it is inserted into the energy
formula it is found that all cycles contribute a negative energy.
When $r<0$ and $g_B>0$, the sign of $g_B(L)$ is different for even
and odd cycles; even cycles contribute negative energy, and odd
cycles contribute positive energy. Cases with $g_B<0$ can be
understood similarly.

\begin{figure}[t]
  \begin{center}
  \centering
  \begin{minipage}[c]{0.30\columnwidth}
      \subfigure[]{ \centering
      \begin{minipage}[c]{0.8cm}
        \centering \includegraphics[scale=0.27]{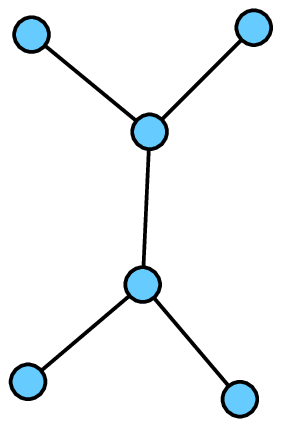}
      \end{minipage}\qquad\qquad\qquad
      \begin{minipage}[c]{0.8cm}
          \centering \includegraphics[scale=0.27]{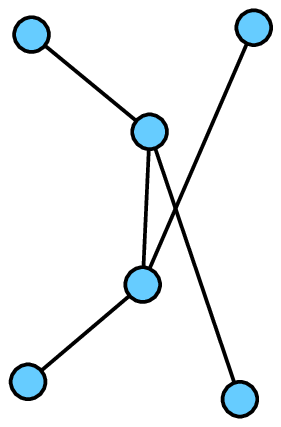}
      \end{minipage}
        }
  \end{minipage}
  \qquad \centering
  \begin{minipage}[c]{0.30\columnwidth}
      \subfigure[]{ \centering
      \begin{minipage}[c]{0.9cm}
          \centering\includegraphics[scale=0.27]{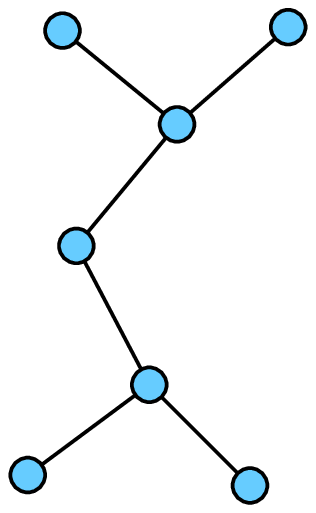}
      \end{minipage}\qquad\qquad
      \begin{minipage}[c]{0.9cm}
          \centering \includegraphics[scale=0.27]{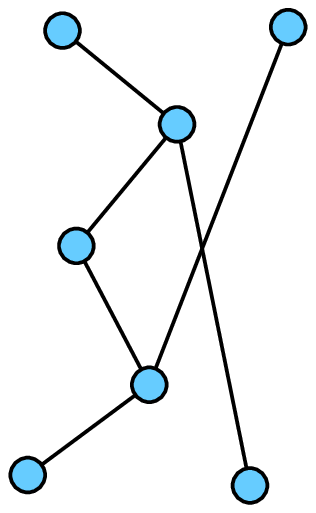}
      \end{minipage}
      }
  \end{minipage}
  \qquad  \centering
    \begin{minipage}[c]{0.30\columnwidth}
      \subfigure[]{ \centering
      \begin{minipage}[c]{1.0cm}
          \centering\includegraphics[scale=0.27]{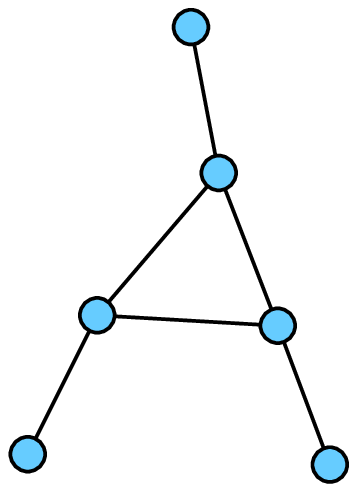}
      \end{minipage}\qquad\qquad\qquad
      \begin{minipage}[c]{1.0cm}
          \centering \includegraphics[scale=0.27]{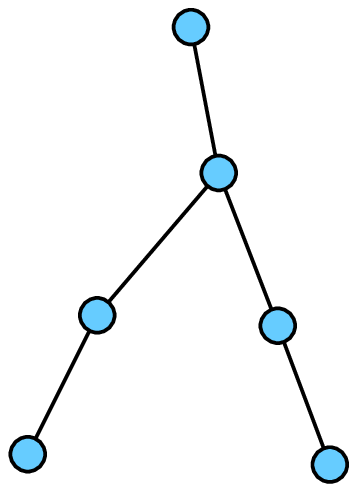}
      \end{minipage}
      }
  \end{minipage}
    \caption{Interaction moves on graphs.}
    \label{fig_moves}
  \end{center}
\end{figure}

Moving on from the definition of the cycle term, note that the
basis states (\ref{Htotbasis}) are all eigenstates of the
operators $H_V$ and $H_B$. In order for the system to evolve from
one graph configuration to another, an interaction term is
necessary. While many forms of interactions acting on graph states
are possible, the ones suggested in \cite{graphity1,graphity2} are
among those that act locally on graphs. Examples of such
interactions are shown in Fig. \ref{fig_moves}. The first two have
the property that they preserve the degree of all vertices; the
third move allows the degree to change. All of them preserve the
connectedness of a graph state.

\subsection{Model with Matter}

Extensions of the basic model may involve introducing a nontrivial
Hilbert space $\HH_{vertices}$ for the vertices or extending the
Hilbert space $\HH_{edge}$ associated with the edges. An example
of the latter, described in \cite{graphity1,graphity2}, consists
of splitting the $|1\rangle$ state in the basic model into three
distinct states so that \be \label{newHHedge} \HH_{edge} =
\mathrm{span}\{ |\,0\rangle, \, |\,1_1\rangle, \, |\,1_2\rangle,
\, |\,1_3\rangle \}. \ee The states $|1_i\rangle$ denote ``on''
edges with different internal labels.

A Hamiltonian for the extended model can depend on the internal
degrees of freedom of the on edges. Thus, in addition to analogs
of the terms described in (\ref{HVB}), other terms are also
possible giving a total Hamiltonian of the from \be
\label{Htotwmatter} H = H_V + H_B + H_{int} + H_M + H_G. \ee Here
the notation $H_M$ and $H_G$ is meant to suggest that these terms
may give rise to effective matter ($H_M$) and perhaps
gravitational ($H_G$) dynamics. These terms could be implemented,
as originally suggested in \cite{graphity1,graphity2}, using the
string-net mechanism \cite{LevinWen1,LevinWen2,LevinWen3} related
to the Kogut-Susskind formulation of gauge theory \cite{Kogut}, or
perhaps by other means, for example, by adapting the framework of
algebraic quantum gravity to dynamical networks
\cite{Giesel:2006uj}.

The specifics of the implementations of these terms will not be
important for the work in the present paper. However, it might be
expected that a realistic implementation should contain wavelike
excitations such as photons or quantum mechanical particle
wavefunctions in the matter spectrum. This rather weak assumption
will be invoked in Sec. \ref{s_Zparticles}.

\subsection{Statistical Interpretation \label{s_Zbare}}

As a statistical system, a graphity model can be studied by
computing its partition function in the canonical ensemble at a
characteristic temperature $T$. As the configurations of the basic
graphity model correspond to graphs, the partition function can be
written as \be \label{Z1} Z = \sum_G \, e^{-\beta E(G)}, \ee where
$\beta = 1/k_BT$ is the inverse temperature, $k_B$ is the
Boltzmann constant, and $E(G)$ is the graph energy computed using (\ref{EGexpectation}).

In models containing extra degrees of freedom $M$, perhaps
corresponding to matter, the partition function must involve
another summation of those states, \be \label{Zgenmatter} Z =
\sum_G \sum_M \,e^{ -\beta E(G) -\beta E_M(G,M)}. \ee The energies
$E_M$ of the additional degrees of freedom can depend on the
underlying graph configuration $G$, and so this dependence is
explicitly shown in (\ref{Zgenmatter}). When the main subject of
study are graphs rather than the matter degrees of freedom, the
summation over $M$ can be hidden by writing \be \label{Zwmatter1}
Z = \sum_G \, z(G)\, e^{-\beta E(G)} \ee with some function
$z(G)$. It will be useful to note that both (\ref{Z1}) and
(\ref{Zwmatter1}) can be cast into the general form \be
\label{ZwmatterC} Z = {\sum_G} \,e^{-C(G)} \ee if the quantity
$C(G)$ is defined as \be \label{criterion} C(G) = \beta E(G) - \ln
z(G). \ee This quantity appearing in the exponent in the Boltzmann
factor can be called the Boltzmann criterion. It reduces to the
usual $\beta E(G)$ if the partition function for matter $z(G)$ is
set to unity.


\section{Statistical Mechanics of the Basic Model \label{s_statmech1}}

The purpose of this section is to estimate the partition function
of a graphity model and use it to understand the statistical
mechanics and ``phase'' structure of the model. Attention is
restricted to classical graphity models so that the sum in the
partition function is over the basis vectors of $\HH_{total}$
only, i.e. states that correspond to classical graphs.
Furthermore, the matter factor is set to \be z(G) = 1 \ee for
simplicity, effectively putting the partition function in the
basic form (\ref{Z1}).

Since the aim of a graphity model is to describe an emergent
geometry, emphasis is placed on understanding the partition
function in the limit of a large number of vertices, $N\ra
\infty$. Since the models considered all have a finite number of
vertices, this limit should be thought of in the following way.
Consider a family of models each with a different $N$ but with the
same other basic parameters (couplings $g_V$, $g_B$ or parameters
$v_0,$ $r$). The equilibrium properties of each of these models
can be evaluated separately. The continuum limit is understood as
the limit of this series of systems arranged in order of
increasing $N$.

The estimate for the partition function below is very crude and is
based on only a few types of graphs. These are described in Sec.
\ref{s_configurations}. It is argued that these few configurations
are sufficient to extract basic information of the model such as
the approximate transition temperatures. These properties, and the
limit of applicability of the results, are described in Sec.
\ref{s_thermodynamics}.

\subsection{Configurations \label{s_configurations}}

In any statistical system, some configurations have more
importance than others in the partition function and thus in
determining the expectation values of observables. In the context
of graph models this means that some graphs, or classes of graphs,
contribute more than others to $Z$. Understanding a graphity model
can therefore be phrased as the problem of identifying the
dominant classes of graphs and evaluating their contributions to
$Z$.

In this section, it is argued that the important classes of graphs
to consider are random graphs and graphs with lowest energy. Other
types of graphs that are of interest and are also discussed are
latticelike graphs. Thus, the partition function is split into \be
\label{Zsplit} Z = Z_{R} +Z_{L} + Z_{H} +\cdots \ee Each of these
terms represents a sum like (\ref{Z1}) restricted to a special
class of graphs and will be estimated separately. The ellipsis
denotes contributions from other graphs that do not fit into the
mentioned categories.

In what follows the components of $Z$ in (\ref{Zsplit}) are
written for large but finite $N$. This means that often only the
dominant behavior of each term, as a function of $N$ and $\beta$,
is presented. Unless otherwise stated, estimates for $\ln Z$ are
valid up to terms or order $N$, which can be neglected given that
the dominant contributions are usually of order $N\ln N$.
Dependence on the inverse temperature $\beta$ is always shown
explicitly even when it appears subleading as a function of $N$.

\subsubsection{ Low-Energy Graphs\label{s_rlcg}}

A class of graphs that can contribute significantly to the
partition function are those which have very small energy and thus
a very large Boltzmann factor. Since the energy of a graph is
determined by the valence and the cycle structure, minimizing the
energy must take both of these properties into account. Consider
regular graphs with varying degrees $v$. The number of cycles
$P(a,L)$ of length $L$ at each vertex $a$ is bounded from above by
\be P(a,L) < c\, v^{L-1} \ee for some constant $c$. Since the
bound is generous, the value of this constant is not important and
can be set to unity. It follows that the total energy due to
cycles is bounded by \be \sum_{a}\sum_L \, g_B(L) \,P(a,L) < \,
g_B \, N \, \frac{e^{|r|v}}{v}. \ee The absolute value in the
exponent is required for the expression to hold for both negative
and positive $r$. Putting this bound together with the energy
contribution from vertex degrees, one obtains another bound
$E_{0,v}$ for the total energy of regular graphs of degree $v$ \be
E(G)> E_{0,v} = g_V N e^{p(v-v_0)^2} - g_B N \frac{e^{|r|v}}{v}.
\ee The first term is minimum for $v=v_0$ and grows
superexponentially for $v$ different than $v_0$. The magnitude of
the second term increases with $v$ but does so more slowly than
the first term. Thus, it is possible to arrange the parameters so
that the lowest possible energy occurs when the degree is $v_0$.
To do this, note that the energy of a $v_0$-regular graph is
smaller than \be E_{v_0} = g_V N \ee and the energy of a
$v_0+1$-regular graph must be larger than \be E_{v_0+1} = g_V N
e^p - g_B N \frac{e^{|r|(v_0+1)}}{v_0+1}.\ee Setting
$E_{v_0+1}-E_{v_0}>0$ one finds \be \label{gVgB} g_V \left( e^p -
1\right) -g_B \frac{e^{|r|(v_0+1)}}{v_0+1} > 0. \ee This is a
conservative requirement on the couplings $g_V$ and $g_B$ that
guarantees the lowest possible energy to be associated with a
$v_0$-regular graph (as opposed to a $v$-regular graph). When the
inequality is satisfied, it also implies that the smallest energy
graph is exactly regular -- all graphs which break the regularity
condition even at a select number of vertices have energy higher
than the minimal energy regular graph. The condition also ensures
that the energy function is bounded from below in the $N\ra\infty$
limit.

In what follows, it will always be assumed that the couplings
$g_B$ and $g_V$ are such that the minimal energy graph $G_0$ is
$v_0$-regular. The contribution of this graph to the partition
function may be estimated as \be Z_L \sim e^{N\ln N} e^{-\beta
E(G_0)}. \ee It will be convenient to write this and all the
following estimates in the form \be \label{ZLestimate} \ln Z_L
\sim \, N\ln N -\beta E(G_0). \ee Here, $E(G_0)$ is the minimum
value of the energy associated to $G_0$. The factor involving
$N\ln N$ is due to the fact that there are on the order of $N!$
ways of assigning labels to the vertices of a graph with $N$
vertices; although these different assignments are sometimes
considered as equivalent from the graph-theoretic perspective,
they correspond to distinct states in $\HH_{tot}$ and thus must be
counted in $Z_L$.

The energy of a graph $E(G_0)$ can also be written as \be E(G_0) =
N (g_V+\epsilon_0) \ee giving $\epsilon_0$ the interpretation as
the average cycle energy per vertex in $G_0$. Because of the form
of the cycle term in the Hamiltonian, this quantity can be (and in
cases of interest, is) negative. With this notation, $Z_L$ becomes
\be \label{Zlowenergy} \ln Z_L \sim N\ln N -\beta N \left(g_V +
\epsilon_0\right). \ee

There may be several other graphs $G$ that have an energy $E(G)$
very close to $E(G_0)$. Such graphs could be interpreted as
perturbations of $G_0$ and would contribute a quantity close to
(\ref{Zlowenergy}) to the full partition function. This could be
taken into account by multiplying $Z_L$ by the number of such
perturbation graphs. However, this multiplicative factor should be
expected to be on the order of $N^x$ for some value $x$ that does
not strongly depend on $N$ or $\beta$. Thus, for large $N$, the
effect of such a multiplicative factor will be negligible compared
with the other factors in (\ref{Zlowenergy}) and hence will not be
included in the analysis.

\subsubsection{Random Graphs\label{s_rrg}}

Random graphs are configurations formed by randomly assigning
edges to a set of vertices \cite{Wormald}. Regular random graphs
are those random graphs that are subject to the constraint that
all the vertices have equal degree. They are widely studied in
graph theory, and a large number of their properties are known.
For example, a typical random graph with a large number of
vertices and low degree has almost no short cycles \cite{Wormald}.
Thus, a typical random graph $G_{R,v}$ with degree $v$ is assigned
almost zero energy by the Hamiltonian $H_B$ so that the total
energy is entirely determined by the valence term \be \label{EGR}
E(G_{R,v}) \sim g_V N e^{p(v-v_0)^2}. \ee Thus, compared to graphs
with low energy discussed previously, the Boltzmann factor of
random graphs can be small. However, because there are many random
graphs in the space of all possible regular graphs, they may
contribute significantly to the partition function overall.

The number of regular random graphs is difficult to evaluate for
general $v$ and $N$ although some asymptotic formulae are known
\cite{Wormald}. For the present purpose, an upper bound $W(N,v)$
for the number of random graphs will be sufficient. In a
$v$-regular graph with $N$ vertices, there are in total $N(N-1)/2$
edges of which $Nv/2$ are ``on.'' The upper bound $W(N,v)$ may be
estimated by the binomial coefficient \be W(N,v) = \left( \begin{array}{c} Nv/2 \\
N(N-1)/2 \end{array}\right) \ee This estimate overcounts graphs,
because it does not take into account the requirement that the
graph must be regular. When $N=6$ and $v=3$, for example, this
estimates gives $W(6,3)\sim 10^{14}$ whereas there are only two
nonisomorphic three-regular graphs with six vertices, for a total
of $2\cdot 6! = 1440$ graphs. Nevertheless, the estimate becomes
more accurate when $N$ is large. In that regime, $W(N,v)$ can be
expressed in a simpler form by writing out the binomial
coefficient in terms of factorials and using the Stirling
approximation. One finds \be \label{Wupperbound} \ln W(N,v) \sim
\frac{1}{2} \,Nv \ln N +O(N). \ee

The random graphs that minimize (\ref{EGR}) are those with
$v=v_0$, and those will be of primary interest here. Thus, using
the observation (\ref{EGR}) and the upper bound
(\ref{Wupperbound}), the estimate for the contribution to $Z$ from
$v_0$-regular random graphs is \be \label{ZGrandom}
\begin{split} \ln Z_{R,v_0} &\sim \,\ln W(N,v_0) -\beta E(G_{R,v_0}) \\
&=\, \frac{1}{2}Nv_0\ln N -\beta N g_V.
\end{split} \ee

Within the random graph category, there are also graphs which
contain more than $Nv_0/2$ edges. These graphs necessarily have at
least two vertices with degree different than $v_0$. Such graphs
have minimum energy if they can be thought of as consisting of a
$v_0$-regular random graph with an extra set of edges. The number
of such graphs could be estimated by taking $W(N,v_0)$ and
multiplying by $N^{2x}$, where $x$ is the number of additional
edges. For small $x$, the energy of such a graph would be
different from $E(G_{r,v_0})$ by $x g_V (e^p-1)$. Thus, the
contribution to the partition function of such graphs would be
written as \be \label{ZGrandomextra} \ln Z_{R,v_0,x} \sim \, \ln
Z_{R,v_0} + 2x\ln N -x \beta g_V (e^p-1). \ee This formula can
only be a reasonable estimate for small $x$. When $x$ becomes
comparable to $N$, some vertices would have degree grater than
$v_0+1$, and the estimated energy shift used in
(\ref{ZGrandomextra}) would break down. In fact, the energy
penalty for those graphs would grow exponentially, and hence they
would contribute negligibly to $Z$.

Summarizing the discussion on random graphs, their total
contribution to $Z$ is \be \label{ZRseries} Z_R = Z_{R,v_0} +
\sum_x Z_{R,v_0,x} + \cdots \ee with $x$ ranging from $1$ to some
number of order $N$ and the ellipsis denoting further
contributions from random graphs with more edges than $Nv_0/2 +
x$. These contributions can be a priori very significant, but
since the goal of the graphity model is to describe graphs with
low degree, the discussion below in Sec. \ref{s_thermodynamics} is
devoted to finding conditions under which these contributions can
be neglected. The terms shown in (\ref{ZRseries}) will be
sufficient for this purpose.

\subsubsection{Homogeneous Graphs\label{s_rlg}}

Lattice graphs are graphs that are regular and further are highly
homogeneous. Because of this, these graphs are of great interest
from the perspective of using a graphity model to describe an
emergent geometry. However, it is important to note that
latticelike graphs may or may not correspond to graphs with lowest
energy. As such, the contribution $Z_H$ of homogeneous latticelike
graphs should be considered separately from $Z_L$.

The estimated partition function $Z_H$ may be written as \be
\label{ZLatestimate} Z_H \sim {\sum_{G}}^\prime \, e^{N\ln N} \,
e^{-\beta N (g_V + \epsilon_H(G))}. \ee This has the same form as
(\ref{Zlowenergy}) but the additional sum is restricted (as
denoted by the prime) to homogeneous, latticelike graphs $G$ each
of which has a cycle energy per vertex denoted by $\epsilon_H(G)$.
The number of terms in the restricted sum that are large depends
on parameters $v_0$ and $r$, but in general can be taken to be
proportional to $N$ or some power thereof. In comparison with the
dependence on $N$ of the individual terms, however, this scaling
with $N$ is insignificant. Furthermore, since the exponents in the
Boltzmann factor of (\ref{ZLatestimate}) are of the same form as
in (\ref{Zlowenergy}) but the energy per node $\epsilon_H$ is at
most as small as $\epsilon_0$, the homogeneous graph contribution
will in general be much smaller than $Z_L$ for large $N$.

\subsubsection{Other Graphs\label{s_og}}

The Hilbert space $\HH_{total}$ also supports a vast number of
graphs that do not fit into the categories described above. Such
graphs could be $v_0$ regular with some short cycles but not as
many as $G_0$, or almost regular with a few vertices breaking the
valence condition, etc. They can be seen as configurations
bridging the gaps between the random, low-energy, and homogeneous
graphs.

Writing down a partition function for these graphs is as difficult
as writing one down for the full model and will not be attempted
here. However, in certain regimes the contribution of these
intermediate graphs can be ignored. In situations where random
graphs dominate the partition function, the intermediate graphs
are subdominant because they are fewer in number. And in
situations where low-energy graphs dominate, the intermediate
graphs are also subdominant, because they have higher energy. In
Sec. \ref{s_simulations} numerical simulations of the full system
will provide a posteriori justification for ignoring these
intermediate graphs.

\subsection{Scaling and Thermodynamics \label{s_thermodynamics}}

Given the arguments in Sec. \ref{s_configurations}, the partition
function of a graphity model may be approximated by the
contribution due to random graphs and low-energy graphs, \be
\label{Zthermo} Z \simeq Z_R + Z_L. \ee Using (\ref{Zthermo}), one
can apply
standard formulae \beqa U &=& -\frac{\pa}{\pa \beta} \ln Z,\\
\label{specificheat} C &=& -\beta^2 \frac{\pa U}{\pa \beta} =
\beta^2 \frac{\pa^2}{\pa \beta^2}\ln Z \eeqa to find the average
energy $U$ or the specific heat $C$. Since $Z$ has a number of
terms that may dominate in some temperature range, one can expect
a number of maxima in the specific heat that correspond to
transitions between the dominance regions of these terms. The
canonical way of finding the temperatures of these transitions
would be to consider $\pa C/ \pa \beta =0$ and solve for $\beta$.
However, the resulting equations are very involved, and the
solutions can only be written neatly after using some
approximations. So, instead, one can treat the problem in a
different manner.

Consider using only pairs of terms from (\ref{Zthermo}) at a time
to calculate the maxima in the specific heat. In
particular, consider the pairs \begin{subequations} \label{PairsofZ} \beqa Z_{L,v_0} &=& Z_L+Z_{R,v_0}, \\ Z_{L,x} &=& Z_L+Z_{R,v_0,1}, \\
Z_{v_0,x} &=& Z_{R,v_0}+ Z_{R,v_0,1}.\eeqa \end{subequations} Each
of these partition functions has only two terms. Thus, specific
heats derived from them have one maximum each marking transitions
between dominance regimes of the two terms. For example, the
function $Z_{L,v_0}$ has a characteristic inverse temperature that
describes a transition from a $v_0$-regular random graph regime
(low $\beta$) to a low-energy graph regime (large $\beta$). As
another example, the maximum in a specific heat derived from
$Z_{v_0,x}$ describes a transition between graphs that should be
expected to be $v_0$-regular and graphs that have $x$ extra edges.

In the limit $N\ra\infty$, the maxima in the specific heats
corresponding to (\ref{PairsofZ}) diverge. The transition inverse
temperatures are given, in the large $N$ limit, by
\begin{subequations} \label{transitionbeta} \beqa \label{bcprediction} \beta_{L,v_0} &=& - \frac{1}{\epsilon_0} \left( \frac{v_0}{2}-1\right) \ln N \\
 \beta_{L,x} &=& -\frac{1}{\epsilon_0-x\Delta\epsilon_V/N} \left( \frac{v}{2}-1+\frac{2x}{N}\right) \ln N\\
\beta_{v_0,x} &=& \frac{2}{\Delta \epsilon_V} \ln N. \eeqa
\end{subequations} In the latter two equations, the quantity
\be \Delta \epsilon_V = g_V \left(e^p-1\right) \ee is always
positive and can be made as large as desired by adjusting $g_V$.
The quantity $\epsilon_0$, being the energy per vertex coming from
the cycle term, is proportional to the coupling $g_B$, and in
cases of interest is less than zero. At this stage, one can also
note that the same transition temperatures could also be obtained
by simply comparing the expressions for the components $Z_L$ and
$Z_R$ without computing the specific heats and their maxima.

As all the expressions (\ref{transitionbeta}) contain a factor
$\ln N$, the transitions they describe all technically occur at
zero temperature in the $N\ra\infty$ limit. This is perhaps not
surprising as it can be already seen in expressions like
(\ref{ZGrandom}) and (\ref{Zlowenergy}) where the leading terms
proportional to $\beta$ are linear in $N$, while the leading terms
independent of $\beta$ scale as $N\ln N$. In the following, it is
useful to either compare the solutions $\beta$ at finite $N$, or
to divide out by $\ln N$ to consider only the coefficients.

In order to reconstruct the transitions in the full system, one
can compare the solutions (\ref{transitionbeta}) pairwise.
Consider first the pair $\beta_{L,v_0}$ and $\beta_{v_0,x}$. If
the coupling condition (\ref{gVgB}) holds (as is necessary for the
estimate $Z_L$ to be consistent), then \be \beta_{L,v_0} >
\beta_{v_0,x}. \ee This implies that the lowest temperature regime
is characterized by low-energy graphs and that there is a
transition to exactly $v_0$-regular random graphs at an inverse
temperature $\beta_{L,v_0}$. The next transition at a much lower
inverse temperature $\beta_{v_0,x}$ occurs between $v_0$-regular
random graphs and random graphs that are not-quite regular. An
important thing to note is that these transitions can be well
separated when $g_V$ is large. Also, note that $\beta_{v_0,x}$
does not depend on $x$.

Next, consider the pair $\beta_{L,v_0}$ and $\beta_{L,x}$. When
$x$ is order unity (as is assumed) and $N$ is large, these two
quantities become equal. It would be incorrect, however, to
conclude that the transition from low-energy graphs to not-quite
regular random graphs occurs at the same temperature as the
transition to exact random graphs. This is because solution
$\beta_{v,x}$ already implies that the transition to not-quite
regular graphs occurs at much lower values of $\beta$. The
apparent paradox is an artifact of using the simplified function
$Z_{L,x}$, which does not take into account regular random graphs
that form an intermediary stage between irregular graphs and
low-energy graphs. The robust conclusion to draw from comparing
$\beta_{L,v_0}$ and $\beta_{L,x}$ is that low-energy graphs are
stable at $\beta$ higher than both of these values.

Comparing the final pair of solutions and observing that
$\beta_{L,x} > \beta_{v_0,x}$ would suggest that graphs are
$v_0$-regular at the highest temperature, irregular at
intermediary temperature, and low-energy at low temperature. While
this is consistent with the hierarchy of the solutions
(\ref{transitionbeta}), it is in conflict with the previous
interpretation. Deciding between the two interpretations requires
some extra input, and this can be obtained by calculating the
expectation values of the energy around the transition
$\beta_{v_0,x}$. It turns out that the original interpretation of
the transitions with the irregular graphs as the highest energy
stage is the correct one and, again, the apparent contradiction
can be traced to the inappropriate use of the expression for
$Z_{L,x}$.

The transition inverse temperatures (\ref{transitionbeta}) must be
understood as only approximations to those one would find by
considering the better approximation (\ref{Zthermo}) or the full
graphity partition function. Nonetheless, the property that the
transition $\beta$s scale logarithmically with $N$ should be
consistent with what one would find from the full partition
function. The estimate $\beta_{v_0,L}$ for the lowest temperature
transition should also be accurate for large $N$ (as will be
confirmed in the next section). And furthermore, the separation of
the transitions from low-energy graphs to regular random graphs
and from regular random graphs to irregular graphs should also
persist in the full model.

More specific questions regarding the nature of the transitions,
however, cannot be addressed reliably. In particular, while
(\ref{transitionbeta}) are all associated with diverging specific
heats, the presented analysis does not make it clear whether these
divergences persist in the full model and whether they can
actually be associated with phase transitions. One possibility is
that the lowest-energy transition described by $\beta_{L,v_0}$ is
a true phase transition while the other transition $\beta_{v_0,x}$
is not. The first part of this expectation may be justified by the
observation that phase transitions occur in other models in which
there are vastly more high-energy states (random graphs) than
low-energy states (low-energy graphs) \cite{InverseFreezing}.
However, like in other statistical models, the details may depend
on the dimensionality of the graph in the low-energy regime. The
second part of the expectation can be based on the fact that no
such dramatic increase in the number of states is present among
the regular and irregular random graphs. In addition to these
uncertainties, it is also not known whether the quantum version of
the model (one in which superpositions of classical graphs are
allowed as quantum states) would exhibit the said phase
transitions.

\section{Simulations \label{s_simulations}}

The main obstacle in evaluating the partition function of a
graphity model is in enumerating the allowed graph configurations
and in estimating their associated energies. Similar difficulties
with other statistical systems have led to the development of
various numerical approximation techniques. Notably among these
are different types of Monte Carlo simulations \cite{Barkema}.
Such techniques can also be used to address various questions
relevant to a graphity model. For example, simulations can search
for the lowest-energy graph for a given set of parameters $N$,
$v_0$, and $r$ and couplings $g_V$ and $g_B$. They can also
provide estimates for the partition function and the expectation
values of obervables, such as the energy, for these parameters and
couplings. This section describes results obtained through
simulations of classical graphity models \cite{JavaCode}.

Given the expectation, justified in Sec. \ref{s_statmech1}, that
states should be $v_0$-regular at the lowest energies, all the
simulations treat exactly $v_0$-regular graphs only. This greatly
simplifies the simulations and eliminates $g_V$ from the list of
parameters that can be explored. The coupling associated with
cycles $g_B$, thus becomes the only coupling in the basic
Hamiltonian and can thus be normalized to unity, \be g_B =1. \ee

Despite the restriction to $v_0$-regular graphs, the scope of the
simulations must be further restricted for computational reasons.
In particular, since counting cycles requires an algorithm that
runs in exponential time in the length of the cycles, simulations
are practically restricted to small values of $L_{max}$ and thus
of $r$.

The results presented were obtained using the Metropolis
algorithm. The moves used to alter the graph during the simulation
are those shown in Figs. \ref{fig_moves}(a) and Fig.
\ref{fig_moves}(b), selected at random at each time step of the
simulation. For a wide range of temperatures, the Metropolis
algorithm performs well and produces results quickly. However, the
graphity Hamiltonian does in principle produce an energy landscape
that has multiple local minima, and hence the Metropolis algorithm
encounters the danger of getting trapped in one of those local
minima at very low temperature and not exploring the full
configuration space. In this work, this problem is avoided mainly
by using relatively small $N$, avoiding extremely large value of
$\beta$, and repeating suspect simulations multiple times if
needed. In some cases, the simulated tempering technique has also
been tried, but all the results reported below rely only on the
Metropolis method.

\subsection{Basic Model}

When studying the basic model only, the general graphity partition
function (\ref{ZwmatterC}) is restricted using \be z(G) = 1,
\qquad C(G) = \beta E(G). \ee Within this framework, consider
models defined by the parameters $v_0=3$ and $r=\pm 2.5$. The
first parameter $v_0$ restricts the simulation to cubic graphs.
The low value of the second parameter $r$ suggests that only short
cycles contribute significantly to the energy of a graph. However,
since the Hamiltonian depends on cycles of various lengths, the
length cutoff $L_{max}$ must be set higher than the minimal cycle
length. In practice, setting $L_{max}=9$ allows to compute a
graph's energy at the percent level.

\subsubsection{Low-Energy Graphs \label{s_lowenergygraphs}}

\begin{figure}[tbp]
  \begin{center}
    \mbox{
      \subfigure[]{\includegraphics[scale=0.27,angle=90]{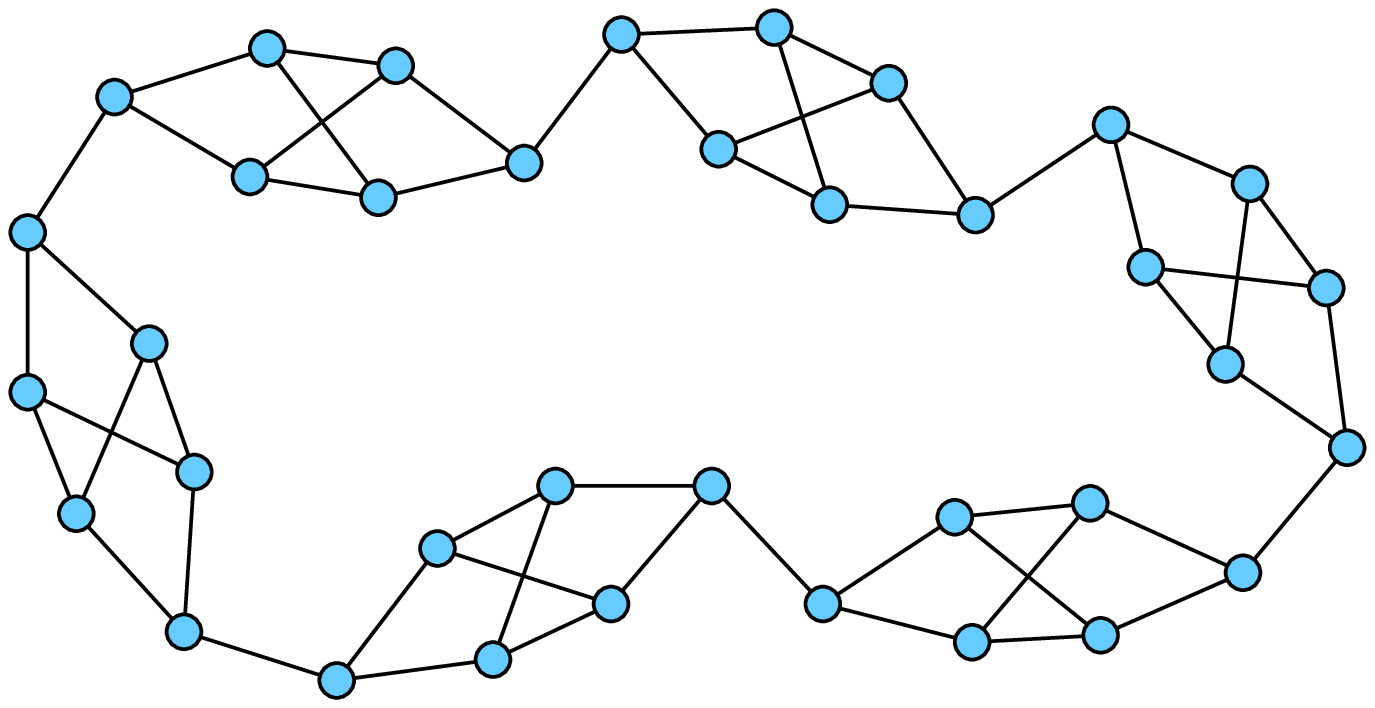}}
      \quad
      \subfigure[]{\includegraphics[scale=0.27,angle=90]{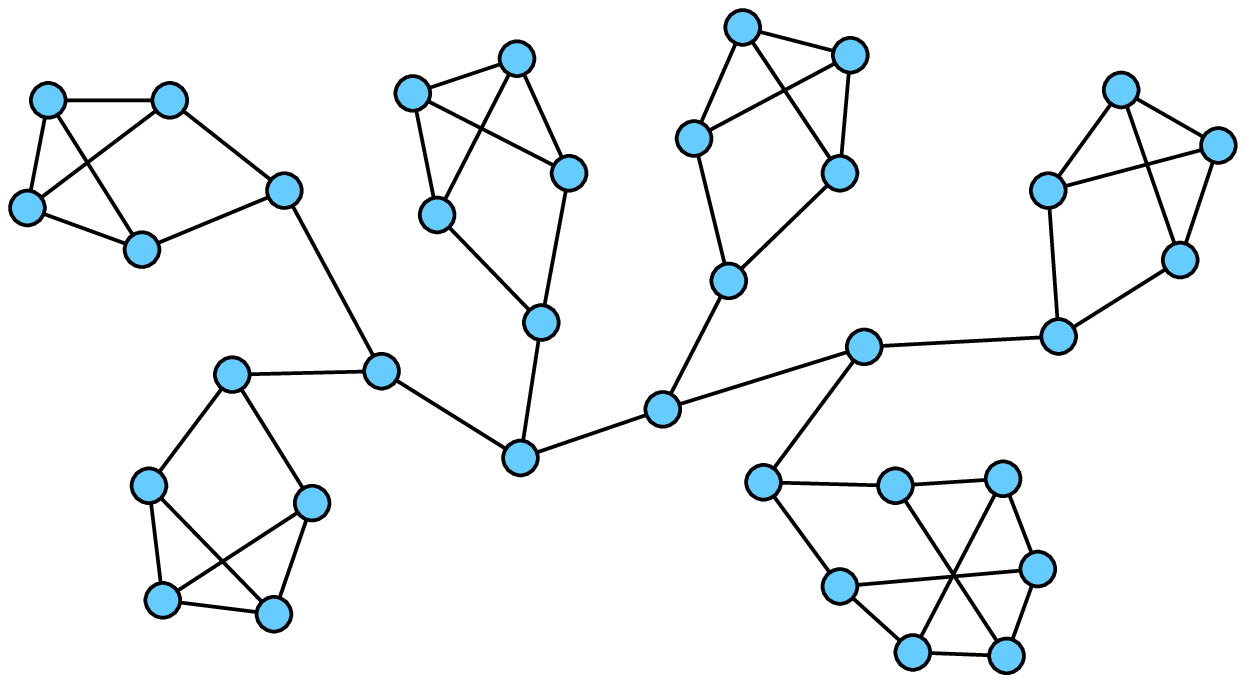}}
      }
    \caption{Lowest-energy graphs in the basic model with $N=36$
    and with (a) $r=-2.5$ and (b) $r=2.5$. The results are obtained through
    simulation at $\beta=0.2$.}
    \label{fig_res25pm}
  \end{center}
\end{figure}

One set of simulations can be performed with the aim of finding
the lowest-energy graph configuration. To do this, the initial
state of the simulation should be sufficiently different from the
expected minimal energy graph; random graphs make for good initial
states and can be generated efficiently \cite{RandomAlgo}. The
simulation should further be performed with $N$ and $\beta$
suitably chosen so that the system evolves rather quickly toward
low-energy states, while still allowing some fluctuations to
occur. Fluctuations ensure that the true, rather than a local,
energy minimum is found. After some trial and error, the values
$N=36$ and $\beta=0.2$ are found to satisfy these criteria. The
lowest-energy graphs found with these settings are shown in Fig.
\ref{fig_res25pm}.

The two configurations shown are very distinct. When $r$ is
negative, the system effectively maximizes the number of 4-cycles
in the graph. This leads to the graph in Fig. \ref{fig_res25pm}(a)
containing six groups of six vertices connected in a chain. For
positive $r$, the system allows some $3$- and $5$-cycles, and this
results in the treelike structure shown in Fig.
\ref{fig_res25pm}(b). Curiously, one of the leaves in Fig.
\ref{fig_res25pm}(b) is different from all the others. This is
because for $N=36$ there can be no graph in which all the leaves
have the most optimal five vertex structure. The position of the
unique leaf on the tree is arbitrary, and thus the lowest-energy
graph is degenerate.

The general features of Fig. \ref{fig_res25pm} persist for other
values of $N$, $v_0,$ and $r$. The lowest-energy states of systems
with negative $r$ are always chains of some small components. The
internal structure and the size of these components depend on the
particular choices of parameters, but the overall one-dimensional
chainlike structure does not. For positive $r$, the lowest-energy
structure is always a tree. Again, the structure of the leaves
varies with $r$.

\subsubsection{Scaling}

The scaling estimates of Sec. \ref{s_thermodynamics} suggest that
systems should become increasingly dominated by random graphs in
the large $N$ limit unless the inverse temperature $\beta$ scales
at least logarithmically with $N$. This effect can be verified
using Monte Carlo simulations.

\begin{figure}[tbp]
  \begin{center}
    \includegraphics[scale=0.65]{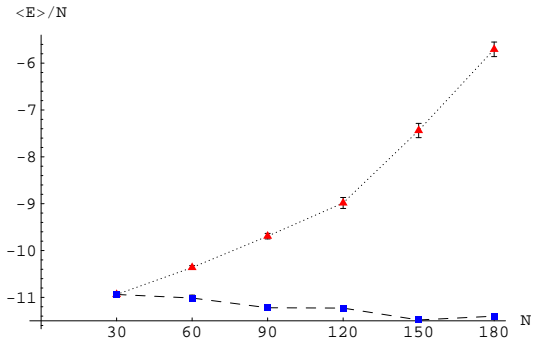}
    \caption{ Average energy per node as function of $N$ in two series of simulations. Data points in the
    rising series are obtained through simulation with $v_0=$, $r=-2.5$, and
    fixed $\beta =0.17$. Data points in the flat series correspond to $\beta$
    scaling logarithmically with $N$. Error bars represent the standard error
    on the mean. The lines shown merely connect the data points and do not represent
    any fit. }
    \label{fig_scaling}
  \end{center}
\end{figure}

Consider performing two series of simulations of the $r=-2.5$
model. In the first series, simulations at different $N$ are
performed always using $\beta=0.17$. In the second series, the
temperatures are chosen according to $\beta=\beta_0 \ln N$ with
$\beta_0 = 0.17/\ln 30$ so that the two series match for $N=30$.
In each simulation of each series, the expectation value (in the
sense of the canonical ensemble) of the cycle energy per node,
$\langle E(G) \rangle / N$, can be computed by extracting a number
of representative configurations from the simulation and averaging
them \cite{Barkema}. Since the lowest energy per node is
independent of $N$ for large N, this expectation value can be
meaningfully compared across simulations with different $N$. The
results are shown in Fig. \ref{fig_scaling}.

In both series, the average energy per node is negative for
$N=30$. This indicates that the system is dominated by
configurations close to the minimal energy state (the lowest
possible energy per node for $r=-2.5$ is $\epsilon_0=-12.2$). In
the series with fixed $\beta$, however, $\langle E \rangle/N$
rises sharply with increasing $N$. This reveals a transition to a
regime of the parameter space where random graphs start dominating
over low-energy graphs. The line does not reach $\langle E \rangle
/N \sim 0$, because for these still small values of $N$ there are
still a large number of graphs that have quite a large number of
cycles but no longer resemble the chain structure. In other words,
points with $\epsilon_0 < \langle E\rangle/N < 0$ represent graphs
of the type omitted in the estimates of the partition function in
Sec. \ref{s_statmech1}.

The second series shown in Fig. \ref{fig_scaling} is the one where
$\beta$ scales with $N$. In this case, the average energy per node
remains very small even for large $N$ indicating that the
low-energy configurations continue to dominate. Note that the dip
at $N=150$, which seems to conflict with the slightly decreasing
trend, is a statistical artifact - the point is actually within
$2\sigma$ of the value corresponding to $N=180$. However, the dip
does indicate that the simulations start becoming less reliable at
high $N$ and increasing $\beta$ because of the ruggedness of the
energy landscape. This unreliability is a systematic error that is
not included in the error bars shown on the plot.

The data in Fig. \ref{fig_scaling}, as all numerical work, cannot
prove the assertion that the required scaling of $\beta$ must
necessarily be logarithmic nor that the logarithmic scaling is
sufficient for the average energy per node to remain close its
minimum. This is particularly so because the explored range of $N$
is small given that the expected dependence of the transition
temperature on $N$ is logarithmic. Nevertheless, the data show the
Monte Carlo simulations of the full model to be consistent with
the expected behavior derived in section \ref{s_statmech1} and
thus provide an important consistency check for those estimates.

\subsubsection{Transitions}

\begin{figure}[t]
  \begin{center}
        \includegraphics[scale=0.65]{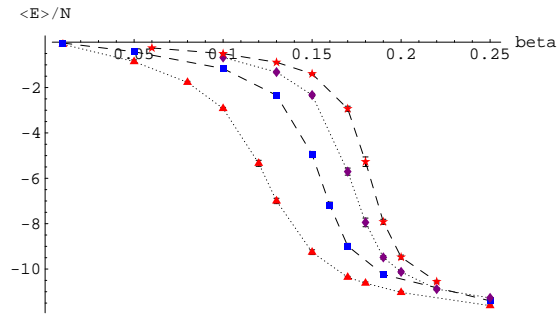}
    \caption{Average energy per node for systems with $r=-2.5$ and $N$
    equal to (left to right) $60,120,180,$ and $240$. Error bars on
    the points (some too small to be seen) represent the standard
    error on the mean. The lines do not represent computed data and
    are only shown to aid the eye distinguish between the four series.}
    \label{f_AvgE}
  \end{center}
\end{figure}

\begin{figure}[t]
  \begin{center}
        \includegraphics[scale=0.65]{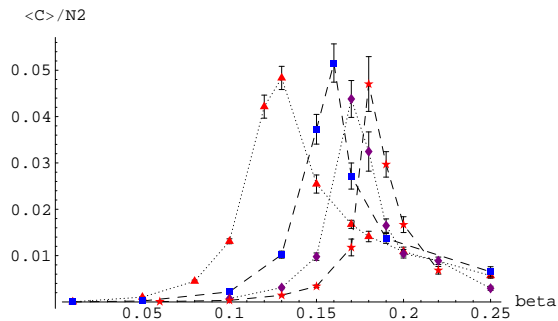}
    \caption{Normalized specific heat for systems with $r=-2.5$
    and $N$ equal to (left to right) $60,120,180,$ and $240$. Error bars
    are computed using the bootstrap resampling method. The lines do not represent computed data.}
    \label{f_SpHeat}
  \end{center}
\end{figure}

Apart from verifying the estimated scaling of $\beta$, one can
also study the low-energy to random graph transition in the
standard manner by performing simulations at various $\beta$ for
fixed $r$ and $N$. The results are shown in Figs. \ref{f_AvgE} and
\ref{f_SpHeat} for systems with $r=-2.5$ and $N=60,120,180,$ and
$240$.

The plot in Fig. \ref{f_AvgE} shows how the average cycle energy
per node decreases as $\beta$ increases. The transition is quite
smooth for $N=60$ but becomes sharper for larger $N$. This effect
is mainly due to the fact that the difference in numbers of states
in the low-energy and random regimes becomes more pronounced with
large $N$. Also, random graphs with $N=60$ might have quite a few
short cycles thus lowering the average energy at low $\beta$. The
location of the transition temperature shifts slightly to larger
$\beta$ with increasing $N$, which is consistent with the scaling
of $\beta$ found in Sec. \ref{s_thermodynamics}.

Figure \ref{f_SpHeat} shows the specific heat associated with the
same simulations. The values plotted are actually specific heats
divided by $N^2$ to make all four curves have roughly equal
height. It should be noted that no special effort was made to find
the true maxima of the specific heats. Doing that would require
more computer time due to the ``critical slowing down'' phenomenon
within the Monte Carlo simulation \cite{Barkema}. However, the
location of these maxima can still be estimated from the plots:
they occur roughly at $\beta=0.13, 0.16,0.17,0.18$. These numbers
can be compared with the estimates from Eq. (\ref{bcprediction}).
Using $\epsilon_0=-12.2$, one finds the expected transition
inverse temperatures to be $\beta=0.17,0.20,0.21,$ and $0.22$. The
estimates therefore consistently overshoot the real transition
values by roughly $20\%$. While the agreement is not stunning, it
does show that the estimate can be a useful guide for determining
where the transition occurs.

\subsection{Matter Using Variance \label{s_simulationsmatter}}

The general partition function of a graphity model, Eq.
(\ref{Zwmatter1}), includes a contribution $z(G)$ from the matter
degrees of freedom. While the form of this matter factor must
depend on how the matter is implemented in the model, it is
reasonable to assume that this factor has some general features.
It may be reasonable to assume, for example, that $z(G)$ is large
for homogeneous graphs and small for graphs that contain many
inhomogeneities. This observation suggests to study the effect of
matter by choosing an ansatz for $z(G)$. A simple one is \be
\label{zGansatz} \ln z(G) = -\mu V(G), \ee with $\mu$ a real
number and \be V(G) = \sum_a \left( E^{\,2}_B(a) -
\frac{E^{\,2}_{B}}{N^2}\right) \ee the variance among the cycle
energies $E_B(a)$ assigned to each vertex $a$ in the graph $G$.
This ansatz implies using the Boltzmann criterion $C(G)$ as \be
C(G) = \beta E(G)+\mu V(G). \ee Since $\mu$ now appears like a
potential in statistical ensembles, it can be called the variance
potential. When it is positive, inhomogeneous graphs are
suppressed in the partition function.

The variance ansatz for $z(G)$ is not intended to mimic a
particular type of matter. Rather, its main advantage is that it
can be implemented in a straightforward fashion in Monte Carlo
simulations. Searches for the lowest-criterion graph can now be
conducted similarly as in Sec. \ref{s_lowenergygraphs} for varying
values of the variance potential $\mu$.

\subsubsection{Cubic Graphs}

\begin{figure}[tbp]
  \begin{center}
    \mbox{
      \subfigure[]{\includegraphics[scale=0.27,angle=90]{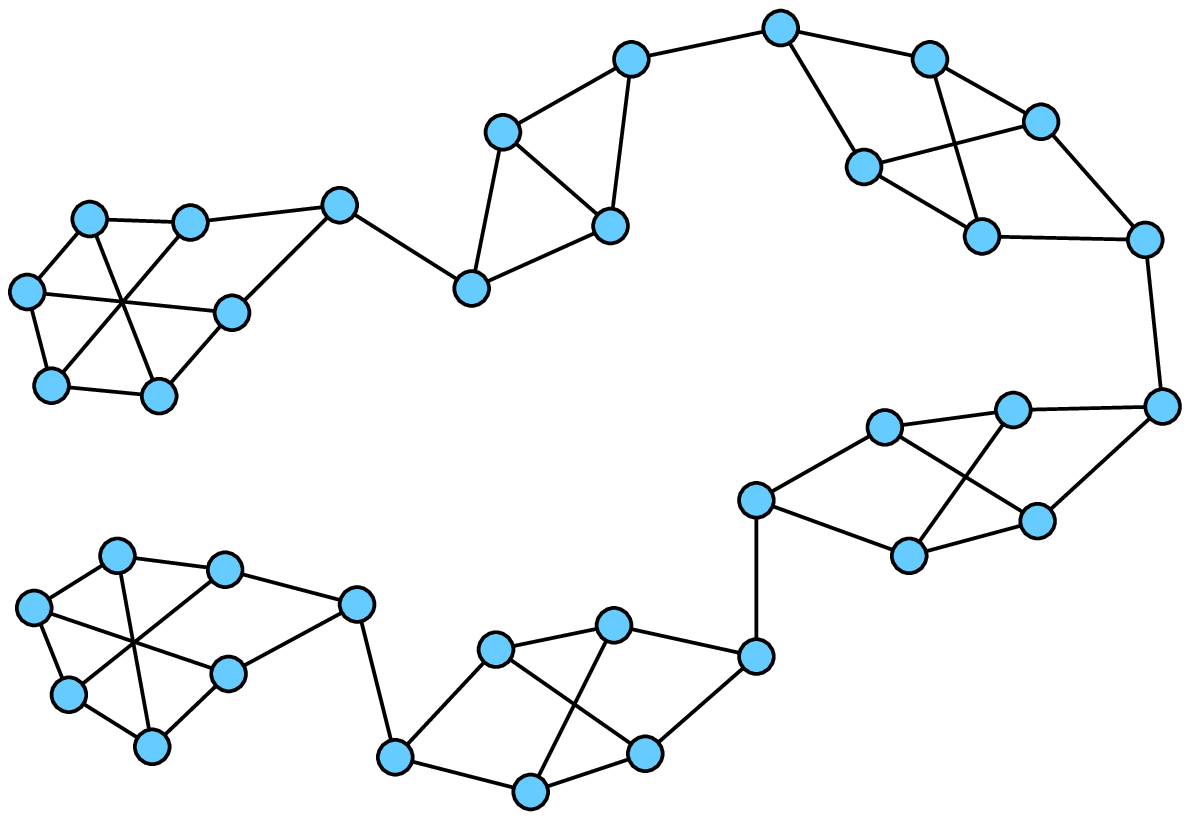}}
      \,
      \subfigure[]{\includegraphics[scale=0.27,angle=90]{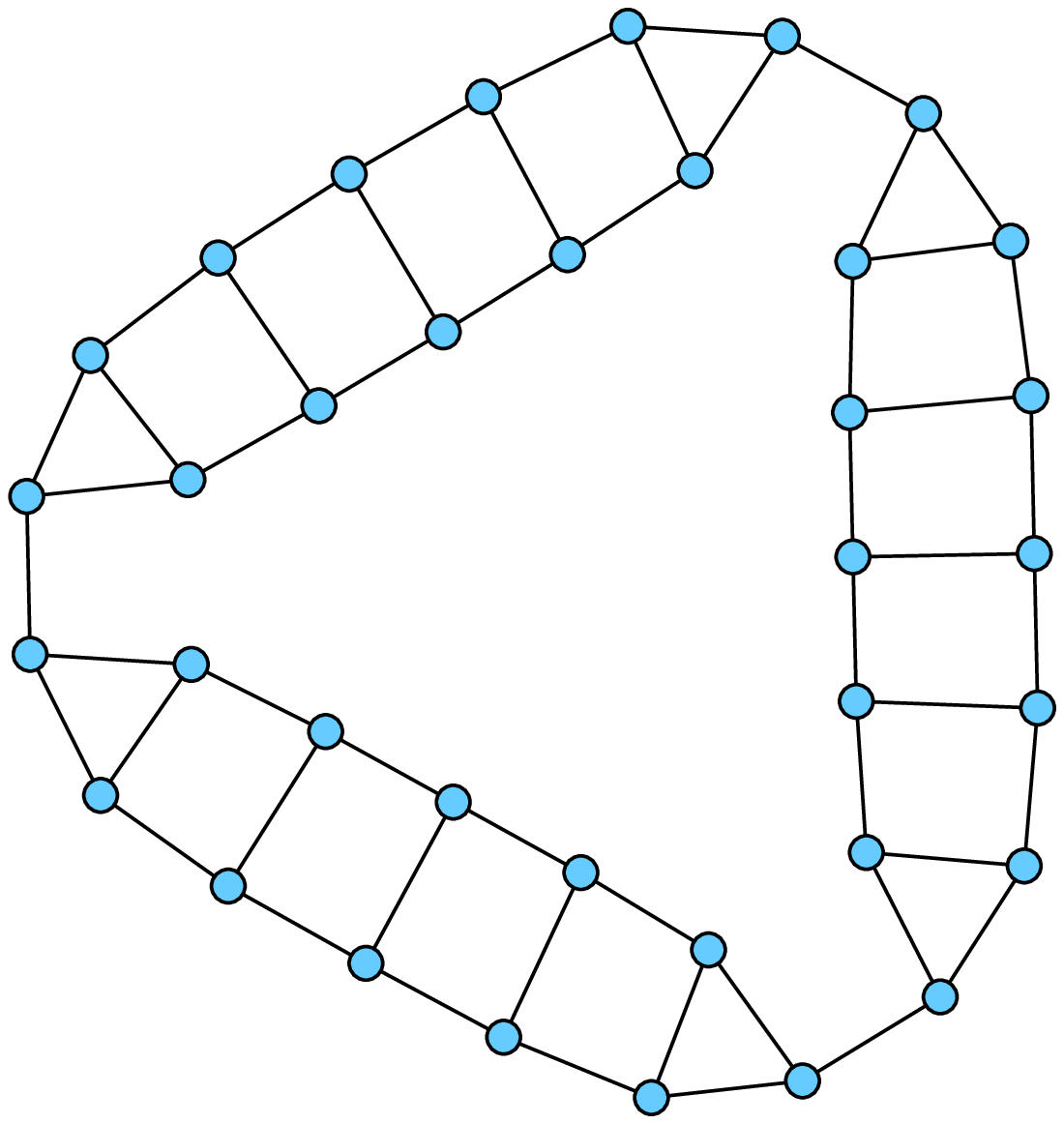}}
      \,
      \subfigure[]{\includegraphics[scale=0.27,angle=90]{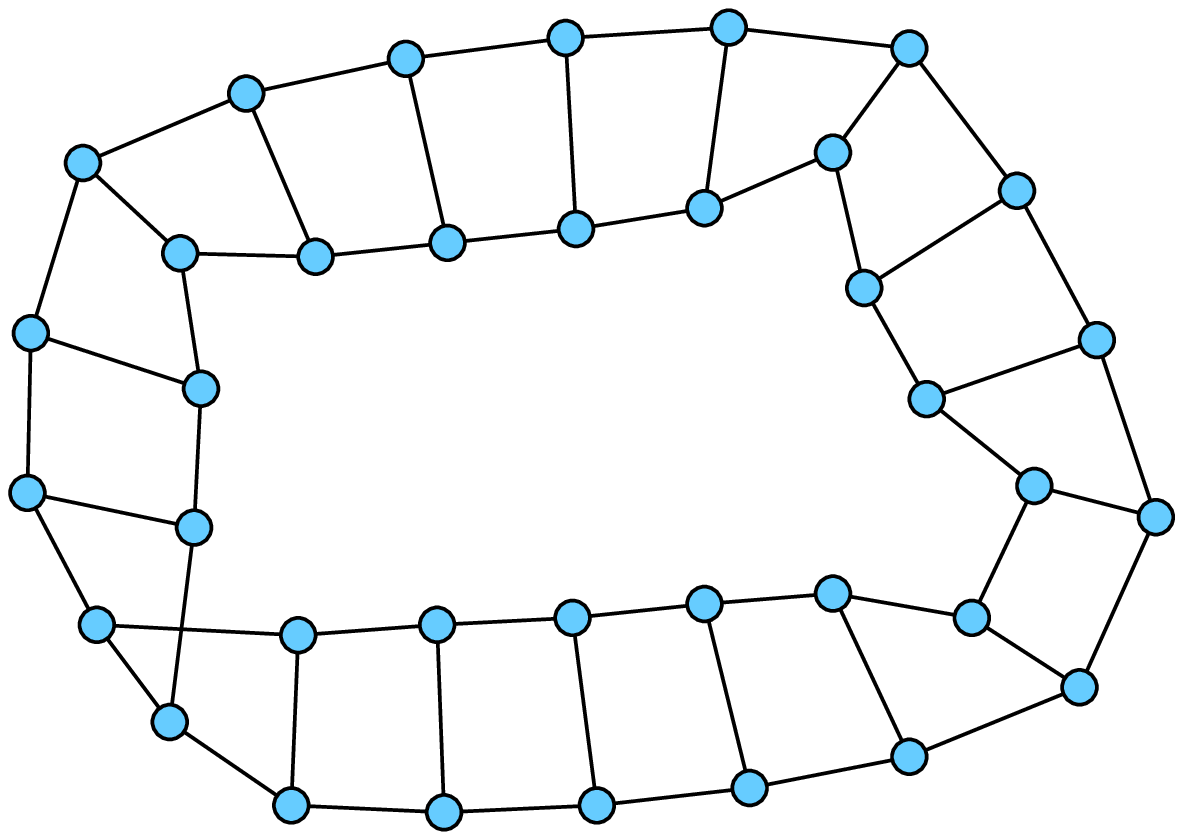}}
      }
    \caption{Graphs with lowest criterion $C$ for $N=36$ and $r=+2.5$.
    Results are obtained with $\beta=0.2$ and the variance potential set
    to (a) $\mu=0.8$ (b) $\mu=3$, and (c) $\mu=4$.}
    \label{fig_res25mu}
  \end{center}
\end{figure}

Consider augmenting the simulations from Sec.
\ref{s_lowenergygraphs} ($N=36$, $v_0=3,$ and $r=\pm 2.5$) by the
matter ansatz with variance potential $\mu$. The results for
$r=+2.5$ are shown in Fig. \ref{fig_res25mu}. The graphs shown are
more and more homogeneous for larger $\mu$, and the ladder graph
shown in Fig. \ref{fig_res25mu}(c) is entirely vertex-transitive.
Interestingly, this graph contains no cycles of length three in
sharp contrast to the ground state graph obtained when $\mu=0$
shown in Figure \ref{fig_res25pm}(b). Simulations of the system
with $r=-2.5$ with large $\mu$ yield the same ladder graph.

In terms of emergence of geometry, the results obtained show that
an effective homogeneous one-dimensional space can be formed from
the graphity model with $r=-2.5$. Moreover, for either sign of $r$
but high enough $\mu$, the emergent space can be the same.

\begin{figure}[tbp]
  \begin{center}
    \mbox{
      \subfigure[]{\includegraphics[scale=0.27]{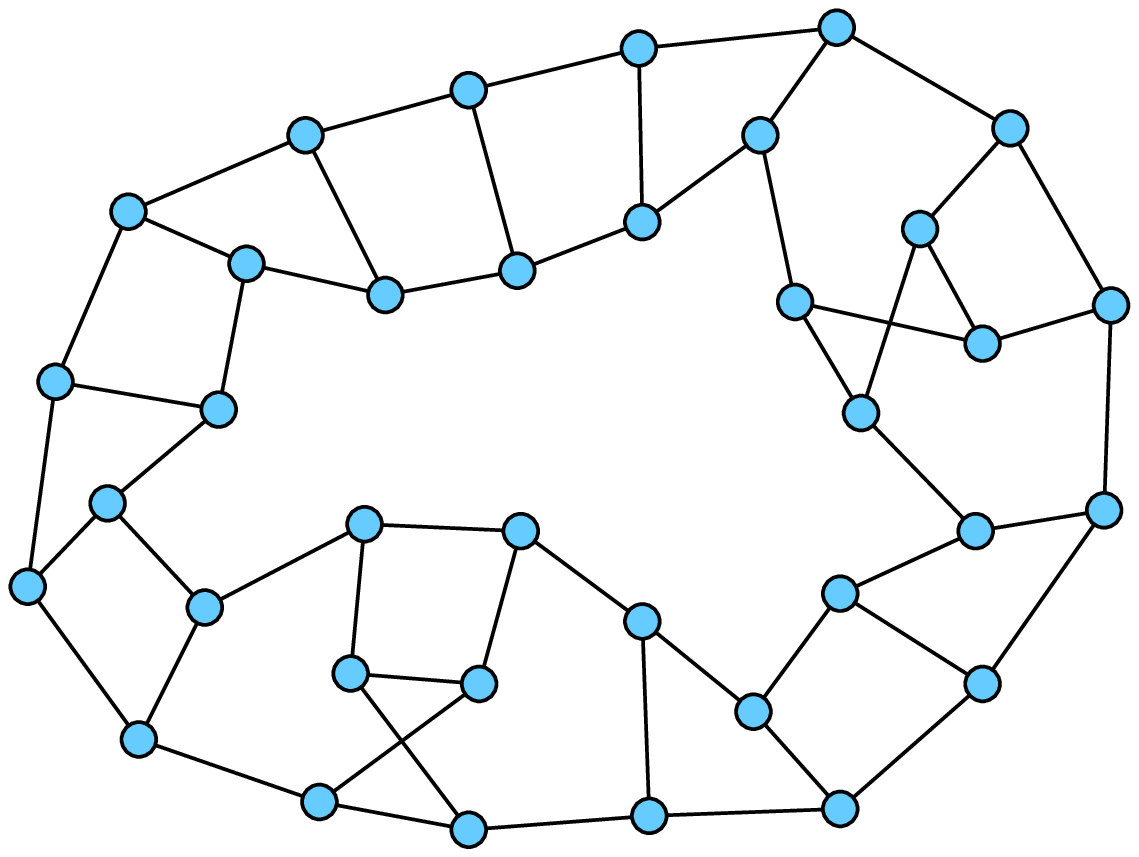}}
      \quad
      \subfigure[]{\includegraphics[scale=0.27]{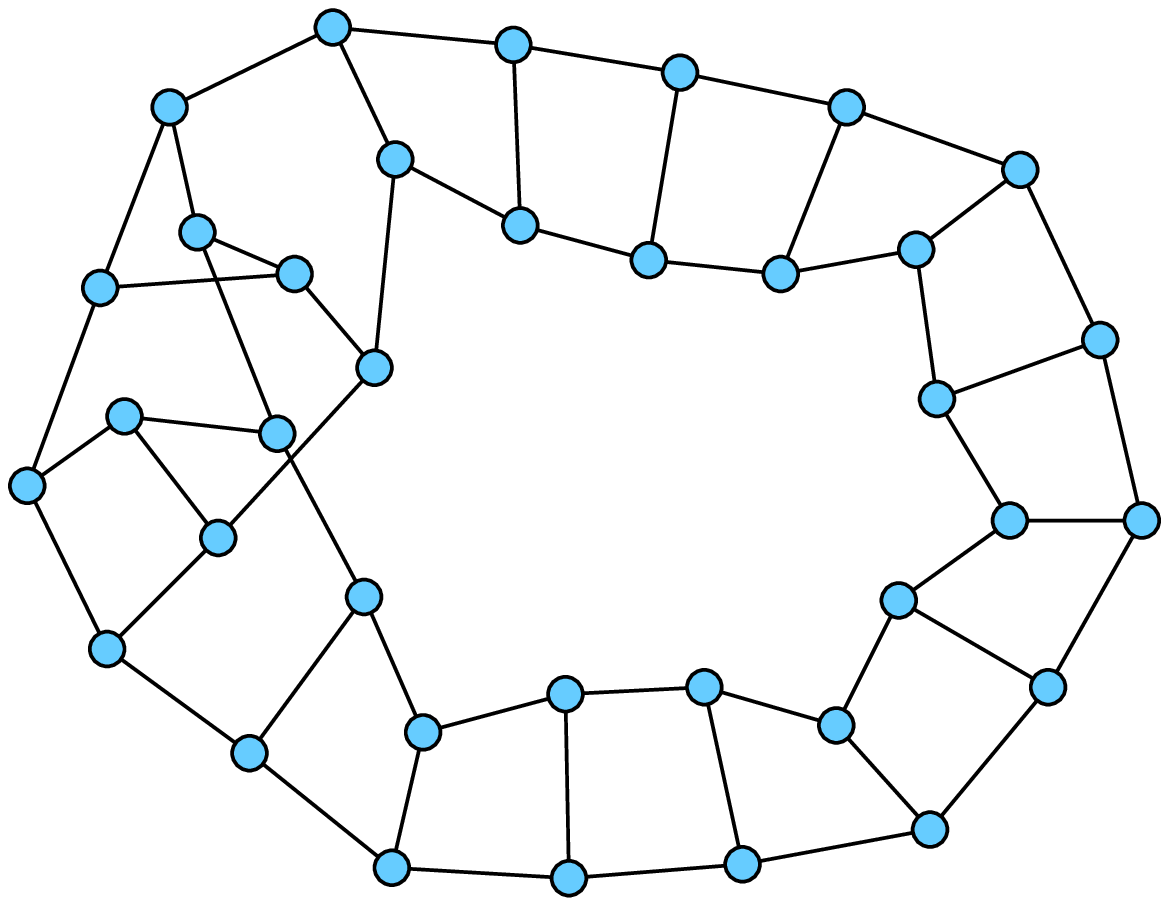}}
      }
    \caption{ Two graphs representing excitations of the $r=-2.5$ and high $\mu$ model.}
    \label{fig_reslikely}
  \end{center}
\end{figure}

Since the simulations leading to Fig. \ref{fig_res25mu} are
performed at low temperature, they can also give some indication
as to what the next-to-lowest-energy graphs look like. Some of
these graph excitations for the $r=-2.5$ simulation are shown in
Fig. \ref{fig_reslikely}. These excitations quickly start looking
like randomized graphs. Nonetheless, one can still recognize the
background lowest-energy graph from their diagrams.

\subsubsection{Quartic Graphs}

Quartic graphs are graphs in which all vertices have degree four.
Simulations with $v_0=4$ can be performed in the same manner as
for cubic case. One new property of quartic graphs (and generally
of regular graphs of degree larger than 3) is that long cycles may
visit a given vertex more than once.

Quartic graphs are of interest because they can give rise to
several types of latticelike configurations. In particular, the
square lattice corresponds to flat 2D space and has four-sided
plaquettes. Also, the diamond lattice corresponds to flat 3D
space, and its shortest cycles have length six.

\begin{figure}[tbp]
  \begin{center}
    \mbox{
      \subfigure[]{\includegraphics[scale=0.27]{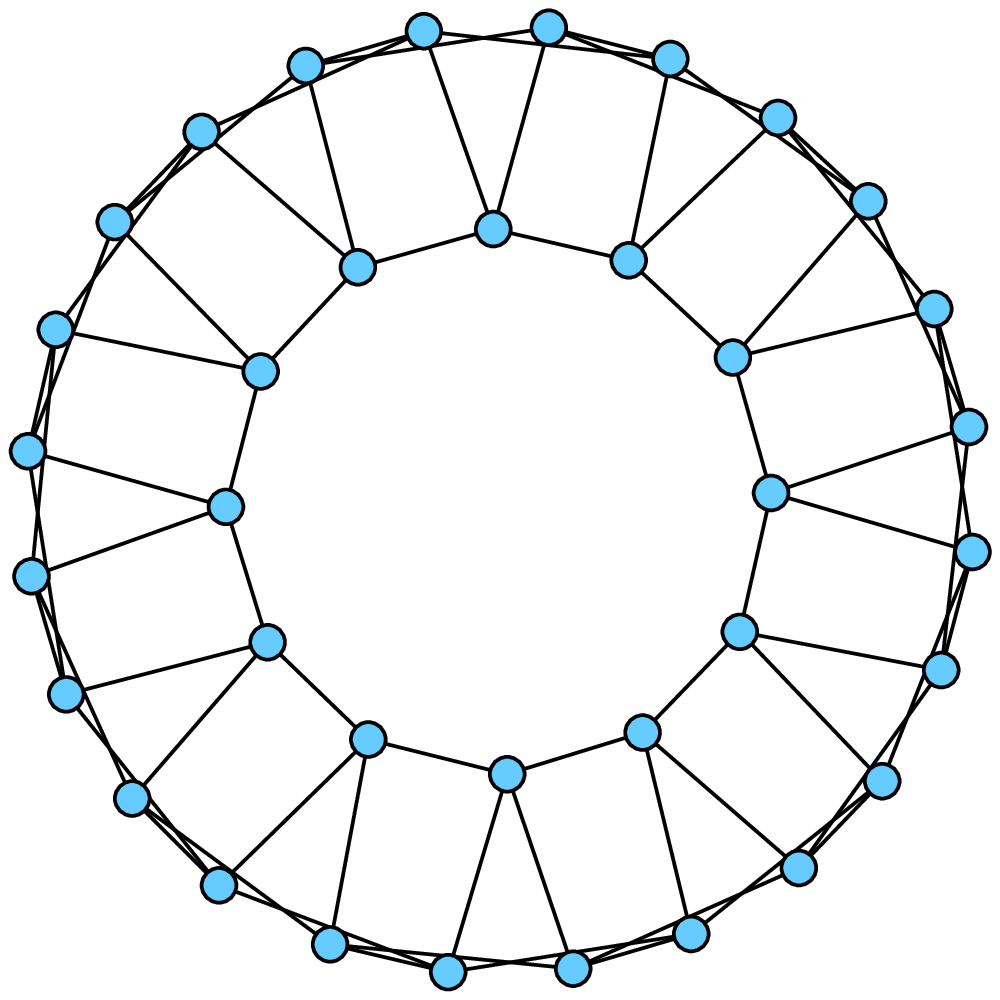}}
      \quad
      \subfigure[]{\includegraphics[scale=0.27]{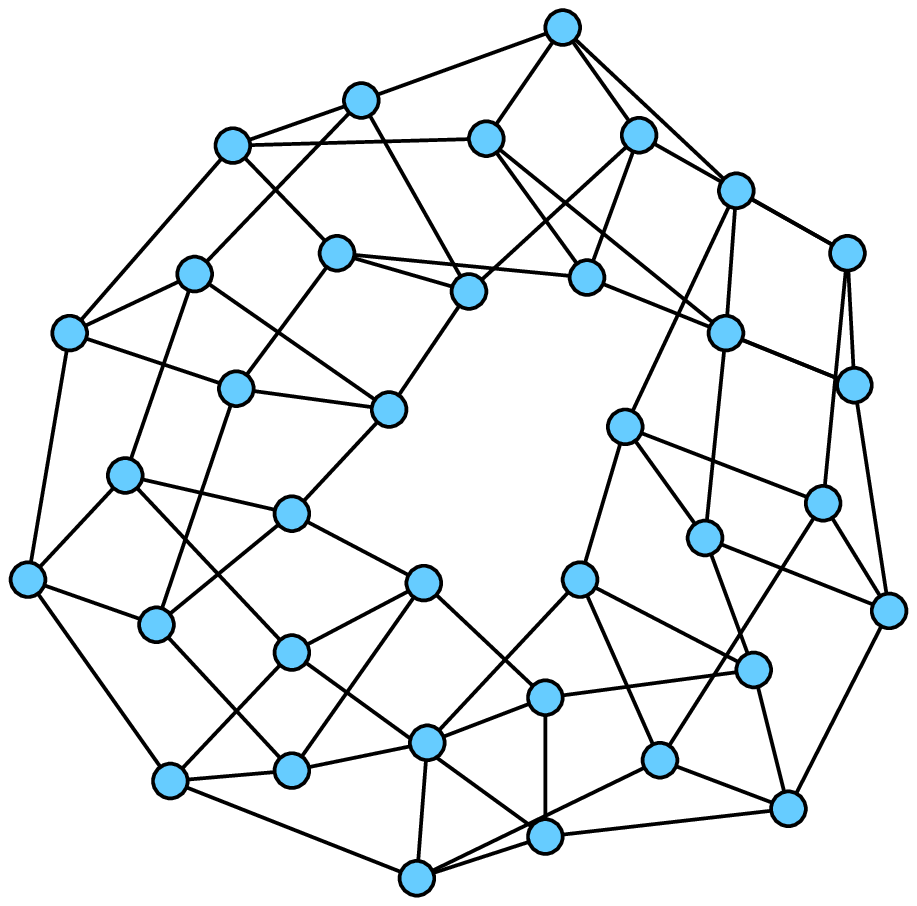}}
      }
    \caption{ Low-criterion quartic graphs.
    (a) is obtained with $r=-1.5$, $\beta=0.6$ and $\mu=32$.
    (b) is produced with $r=-2.5$, $\beta=0.17$, and $\mu=3.8$. }
    \label{fig_quartic}
  \end{center}
\end{figure}

Simulations searching for graphs with lowest criterion $C(G)$ were
performed for small values of $r$ and large variance potential
$\mu$. Some of the results are shown in Fig. \ref{fig_quartic}.
Neither of the shown graphs is planar and this indicates that the
emerging graphs might be interpreted as having dimension greater
than one. Indeed both graphs in Fig. \ref{fig_quartic} may be
thought of as two-dimensional tori. This is particularly
convincing in the case of Fig. \ref{fig_quartic}(b), where the two
circumferences associated with a torus have lengths four and nine.

\subsubsection{Graphs with Larger Plaquettes}

Simulations with $|r|$ much greater than $2.5$ encounter at least
three problems. First, as $|r|$ increases, the evaluation of the
graph energy becomes more time-consuming as longer cycles must be
taken into account. Second, in order to obtain results that can be
generalized to the continuum limit, the number of vertices $N$
used in the simulation must be large enough so that in the
high-$\beta$ regime, the energy per node of the emerging graph is
not skewed by the finite value of $N$. And third, as $|r|$
increases, the depth of the valleys in the energy landscape
increases. All three of these issues make it more difficult for a
simple desktop simulation to provide accurate results quickly.
More sophisticated methods like simulated tempering should likely
be used to make real progress.

It is possible to perform some quick simulations with large $|r|$,
for example $r=-6$, if $L_{max}$ is set very small, for example
$L_{max}=7$. Such a low cutoff value for the cycle length means
that the evaluation of the energy is not at all accurate. However,
all three problems mentioned above become less serious.
Simulations performed in this way with large $\beta$ and large
$\mu$ produce graphs with hexagonal plaquettes that resemble those
graphs described in \cite{graphity2}. While this is encouraging in
the sense that it suggests that interesting lattices may appear in
other regions of the graphity parameter space, it should be
stressed again that such simulations with low $L_{max}$ cannot
capture properties of the full graphity Hamiltonian. Proper
simulations at large $|r|$ must thus be considered beyond the
scope of this initial numerical exploration.


\section{Statistical Mechanics of a Graphity Model with Particles \label{s_Zparticles}}

In the previous section, the effect of matter on graph
configurations was modeled by adding a term proportional to the
vertex energy variance to the Boltzmann criterion. Simulations
showed that large values of the variance potential can indeed
change the type of graph that minimizes the Boltzmann criterion
(\ref{criterion}) and thus change the structure of the expected
graph at low temperatures. But while the variance prescription is
straightforward to implement in a Monte Carlo simulation, it does
not fully incorporate the effect of matter on a graph. It does
not, for example, account for the fact that spaces with different
geometries may allow for different densities of particle states.
Thus, it would be desirable to supplement, or replace, the
variance prescription with something more realistic.

Since explicitly evaluating the partition function of a graphity
model, with or without matter, is not feasible without
approximations, one cannot, at least at this stage, learn very
much by studying a particular form of matter coupling to graphs.
This section therefore focusses on general properties of wavelike
matter and discusses how they affect a dynamical geometry such as
in a graphity model. The main principles leading to the argument
that matter can significantly alter, and even determine, the shape
of a dynamical geometry are outlined below in section
\ref{s_flexibleboxes}. These principles are then applied to
dynamical graphs in Secs. \ref{s_flexibleconfigurations} and
\ref{s_flexiblescaling}. As in Sec. \ref{s_statmech1}, only the
leading terms in $N$ and $\beta$ will be written in the estimates
for the partition function components.

\subsection{Flexible Boxes \label{s_flexibleboxes}}

Consider a set of $N$ particles in a two-dimensional rectangular
box with fixed area $A$ in which the lengths $L_x$ and $L_y$ of
the two sides can vary as long as $L_x L_y = A$. As the geometry
of the box is not fixed, the partition function of the system
includes a summation over the configurations, denoted $(L_x,L_y)$,
that the box can take. It can be written as \be
\label{Zflexiblebox} Z = \sum_{(L_x,L_y)\in \mc{C}} z(L_x,L_y)
e^{-\beta E_B(L_x,L_y)}\ee where $E_B(L_x,L_y)$ is the energy
associated with a given box, and $z(L_x,L_y)$ describes the matter
content. In the case that $E_B(L_x,L_y)=0$ for all $L_x$ and
$L_y$, the properties of the system, including its geometry, are
determined by the matter factors and the space of configurations
$\mc{C}$ over which the sum is taken.

The space of configurations is the set $\mc{C}$ of possible pairs
of lengths $(L_x,L_y)$. For the present purpose, the precise form
of this set will not be important, and only some of its general
features will be assumed and used. First, the set $\mc{C}$ should
contain certain extreme configurations. It should contain a
configuration very close to $(L_x,L_y)=(\sqrt{A},\sqrt{A})$, i.e.
a square box. It should also contain some configurations that are
as long-and-narrow as possible. Assuming that the shortest allowed
length is $\ell$, these extreme long-and-narrow configurations are
$(\ell,A/\ell)$ and $(A/\ell,\ell)$. Second, the total number of
configurations in the set $\mc{C}$ should scale as some power of
$A/\ell^2$. This condition can be implemented if the lengths $L_x$
and $L_y$ are discretized. It should be noted that while the
minimal length scale $\ell$ appears in several of the formulae
below as a regularization parameter, the limit $\ell\ra 0$ can be
taken at the end without altering the conclusion of the main
argument.

For the matter factors, consider $z(L_x,L_y)$ to be partition
functions for a collection of $N$ particles. Thus \be
\label{zNbox} z(L_x,L_y) =
\frac{1}{N!}\left(z_1(L_x,L_y)\right)^N\ee with \be z_1(L_x,L_y) =
\sum_{E_{L_x,L_y}} \!\! e^{-\beta E_{L_x,L_y} }.\ee The summation
in $z_1$ is over the energy states available to a single particle
in a box with side lengths $L_x$ and $L_y$. These states may
correspond to the standing waves that can form inside the
two-dimensional box and, if the matter is quantum mechanical,
their energies may be associated with inverse wavelengths. For
nonrelativistic particles with mass $m$, the energy states
$E_{L_x,L_y}$ could be given by \be E_{L_x,L_y} = \frac{1}{8m}
\left( \frac{n_x^2}{L_x^2} +\frac{n_y^2}{L_y^2}\right) \ee with
$n_{x}$ and $n_y$ the usual integer quantum numbers.

It is instructive to look at the lowest-energy particle states
available in the extreme box configurations. In the large area
limit these are \be E_{min }= \left\{ \begin{array}{ll}
(8m\ell^2)^{-1} &\quad{\mbox{for }}\, (\ell,A/\ell) \mbox{ and }(A/\ell,\ell) \\
0&\quad {\mbox{for }}\, (\sqrt{A},\sqrt{A}). \end{array} \right.
\ee The discrepancy between these two energies may be very large
when $\ell$ is small. Minimal particle energies in nonextreme
configurations lie between the two shown values.

With this particular choice of particle energy spectrum, the
matter partition functions become \be \label{littlez1}
z_1(L_x,L_y) =\sum_{n_x} \mathrm{exp}\!\left( -\frac{\beta
\,n_x^2}{8mL_x^2}\right)\; \sum_{n_y} \mathrm{exp}\!\left(
-\frac{\beta \,n_y^2}{8mL_y^2}\right). \ee While exact expressions
for (\ref{littlez1}) are not known, they can be approximated by
the number of states for which $\beta$ times the particle energy
is less than unity\footnote{In the special case that $L_x=L_y$,
$z_1$ may perhaps be better estimated using an integral expression
\be \nonumber z_1 \sim \int_0^\infty e^{-\widetilde{\beta} n^2}
\!n \; dn\ee with $\widetilde{\beta}$ being a rescaled version of
$\beta$. This approach, however, does not significantly differ
with the other scheme.}. Such a prescription should at least
provide the leading behavior of these factors. One finds \be
\label{littlezscaling} z_1\left(L_x,L_y \right) \sim \left\{
\begin{array}{ll}  \!A \,
\mathrm{exp}\left(\frac{-\beta}{8m\ell^2}\right)
\sqrt{\frac{8m}{\beta\ell^2}}
&\quad{\mbox{for }}\, (\ell,A/\ell) \mbox{ and } (A/\ell,\ell) \\
\!A \, \frac{8m}{\beta} &\quad {\mbox{for }}\,
(\sqrt{A},\sqrt{A}).
\end{array} \right. \ee
In the first line, it was assumed that $\beta/(8m\ell^2)>1.$

While the leading behavior with $A$ is the same in both cases, the
contribution from the square box can be larger than those from the
long-and-thin boxes when $\ell$ and $m$ are small and $\beta$ is
large. In such cases, the dominance of the square configuration
will be further amplified when $z_1(L_x,L_y)$ is raised to the
power of $N$ in (\ref{zNbox}). Of particular interest is taking
$N$ to scale with some power of $A$, i.e. $N\sim c A^\alpha$; the
special case $\alpha=1$ corresponds to considering a set of boxes
of various areas $A$ with fixed particle density. In the case with
$N\sim A$, the matter factor $z(L_x,L_y)$ will be exponentially
larger for the square configuration than for any of the other
long-and-thin configurations. Moreover, since the number of
nonsquare configuration scales as a power of $A/\ell^2$, the
contribution of the square configuration will dominate the
partition function for this toy model in the limit $A\ra\infty$.
Similar scaling arguments can be used to show that expectation
values of observables and their fluctuations will be determined
primarily by their values on the square configuration.

This example is an example of how matter degrees of freedom can
determine the shape of a dynamical geometry. The result is quite
striking -- out of the large set of configurations $\mc{C}$,
matter can select a special configuration and effectively force
all observable physics to be based on this special geometry.

\subsection{Configurations \label{s_flexibleconfigurations}}

The discussion of particles in a flexible box may be adapted to
the case where the trapping container is a graph. The purpose of
this section is to estimate the matter factors $z(G)$ associated
with different classes of graphs in the same spirit as
$z_1(L_x,L_y)$ and $z(L_x,L_y)$ were estimated above. In all
cases, the number of particles, denoted $N$ above, will be assumed
to be the same as the number of vertices present in the graph.
Thus one of the key assumptions in the following analysis is that
the density of particles is constant across graphs with different
number of vertices $N$.

\subsubsection{Homogeneous Graphs}

Homogeneous graphs are the easiest ones to deal with since their
large-scale properties resemble lattices. Because the graphity
models all have a finite number of vertices, these homogeneous
lattices are closed geometries and, as discussed in Sec.
\ref{s_simulationsmatter}, represent two-dimensional tori and
similar compact structures in other dimensions.

Consider first two-dimensional tori. They can be described by two
lengths $C_1$ and $C_2$ representing the two circumferences. The
problem of evaluating $z(G)$ for such graphs is in direct
correspondence to evaluating $z(L_x, L_y)$ using (\ref{littlez1})
and (\ref{zNbox}) above. One modification that needs to be made
lies in replacing the area by the number of vertices, \be A = N
\ell^2; \ee the factor $\ell^2$ is required for dimensional
reasons. Also, since each particle's wavefunction as well as its
derivative should be continuous on the whole torus, the summation
over particle energy states should be restricted to even integers.
This restriction only rescales the estimates
(\ref{littlezscaling}) by some factor of order one and thus the
estimates made in the present case are equivalent to
(\ref{littlezscaling}).

To obtain the full contribution of the lattice graphs to the
partition function, the estimates for $z(G)$ must be combined with
the nonmatter components of (\ref{ZLatestimate}). For a torus with
two circumferences almost equal, $C_1\sim C_2\sim \ell\sqrt{N}$,
one obtains \be \label{ZHfattorus} \ln Z_{H,\, C \sim
\ell\sqrt{N}}^{\,2d} \sim \, N\ln N -\beta N
\left(g_V+\epsilon_H\right) + N \ln \frac{8m\ell^2}{\beta}.\ee For
a long-and-thin torus, $C_1\ll C_2$ or $C_2\ll C_1$, one has \be
\label{ZHthintorus} \ln Z_{H,\, C_1\ll C_2}^{\,2d} \sim N\ln
N-\beta N \left(g_V+ a \epsilon_H +\frac{1}{8m\ell^2}\right)
+\frac{1}{2}N \ln \frac{8m\ell^2}{\beta}. \ee In both cases, note
that the factor $A^N= (N\ell^2)^N$ appearing after raising $z_1$
to the power of $N$ is cancelled by the $N!$ in the denominator of
(\ref{zNbox}). In the equation for the long-and-thin torus, the
factor $a$ appearing in front of $\epsilon_H$ is a real number of
order one that reflects the fact that the energy per node
associated with the two types of tori may be different. In
particular, since the long-and-thin torus contains winding cycles
that are not present in the isotropic torus \cite{graphity2}, this
factor should be expected to be greater than one.

Next, consider homogeneous graphs that represent generalizations
of tori in higher dimensions $d$. For simplicity, consider the
isotropic configurations in which all the circumferences $C \sim
\ell N^{1/d}$. The generalization of (\ref{littlez1}) for this
case can be written as \be z_{1,d}(C\sim \ell N^{1/d}) =
\sum_{\overrightarrow{n}} \mathrm{exp} \left( -\frac{\beta
\overrightarrow{n}^2}{8m N^{2/d}}\right) \ee with
$\overrightarrow{n}$ being a $d$-dimensional vector of quantum
numbers. Using the same approximation as before, that the sum is
approximately equal to the number of states such that the exponent
is less than unity, and assuming that
$\beta(8m\ell^2)^{-1}N^{-2/d}<1$, one obtains\footnote{If the sum
were approximated using the integral approximation in footnote
$1$, this would have an additional factor depending on $d$. This
discrepancy will not be important for the argument below.} \be
z_{1,d}(C\sim \ell N^{1/d}) \sim N \,
\left(\frac{8m\ell^2}{\beta}\right)^{d/2}, \ee i.e. the same
dependence on $N$ as given by the expression
(\ref{littlezscaling}). This is quite important: despite the
number of states on a $d$-dimensional space up to some level $|n|$
growing as $|n|^d$, this growth is largely balanced by the smaller
lengths $L\sim \ell N^{1/d}$, resulting in an overall $z_1$ that
does not depend very strongly on $d$. After using this to evaluate
the full contribution of the $d$-dimensional torus to $Z_H$, one
therefore finds an expression that has almost the same form as
(\ref{ZHfattorus}), namely \be \label{ZHdtorus} \ln Z_{H,\, C\sim
N^{1/d}}^{\,d} \sim
 \, N\ln N-\beta N \left(g_V+a_d \, \epsilon_H\right) + \frac{1}{2} Nd \ln \frac{8m\ell^2}{\beta}.
\ee The most important difference between this expression and
(\ref{ZHfattorus}) lies in the different energies, encoded in the
real numbers $a_d$, associated by the graphity Hamiltonian to the
different tori.

Estimating the contribution of nonisotropic $d$-dimensional tori
to $Z_H$ can be done in a similar fashion. As in the case of
two-dimensional tori, such nonisotropic structures would
contribute terms which, like (\ref{ZHthintorus}), would contain
factors $\mathrm{exp}\left( -\beta N /(8m\ell^2)\right)$. All the
different terms will not be written out here as they will turn out
to not to play an important role.

\subsubsection{Low-Energy Graphs}

Simulations show that the lowest-energy graphs are one-dimensional
chains made up of subgraphs with a small number of vertices. Since
such graphs are not homogeneous lattices, one has to make further
assumptions in order to guess what their matter factors should be.
Suppose that these chain graphs can support particle states as
standing waves both along the one-dimensional chains and also
within the individual subgraphs. Then, since the substructures
only have a finite number of vertices, standing waves supported
within the subgraphs will have wavelengths no longer than a few
$\ell$. Thus, these standing waves will have a large energy; the
situation is not dissimilar than for the long-and-thin boxes or
tori.

This reasoning suggests to write the contribution $Z_L$ with a
factor that penalizes chain graphs for high-energy particle
excitations, \be \label{ZLowmatter} \ln Z_L \sim N\ln N-\beta N
\left(g_V+\epsilon_0+\frac{1}{8m_L \ell^2}\right). \ee Here, $m_L$
is an effective mass corresponding to the particles on the
low-energy graph. It is thought to absorb all the factors that are
missed in this simple estimation. What will be important is that
this effective mass should not scale significantly with $N$ and
should always be positive.

\subsubsection{Random Graphs}

As in the case of low-energy graphs, it is unclear how to estimate
the matter factor $z(G)$ associated with random graphs. In
general, the term for $v_0$-random graphs can be written as \be
\label{ZRwithmatter} \ln Z_{R,v_0} \sim \, \frac{1}{2}Nv_0 \ln N
-\beta N g _V + \ln z(G_R). \ee Since the average distance between
two vertices in a random graph is small \cite{Wormald}, the
largest wavelengths associated with particle states should be on
the order of $\ell$. This would suggest to write this factor
similarly as in (\ref{ZLowmatter}), for example, \be \ln z(G_R)
\sim - \frac{\beta N}{8m_R\ell^2} \ee with some effective positive
mass $m_R$. Such a matter factor would be effectively penalizing
random graphs for their tight connectivity.

A discussion of random nonregular graphs would follow the same
lines. It will not be necessary, however, to include it here.

\subsection{Scaling \label{s_flexiblescaling}}

Armed with estimates for the important components of $Z$, one can
repeat the analysis of Sec. \ref{s_thermodynamics} for the
graphity model with matter. However, given that it was found
previously that basic intuition can be extracted quite easily just
by comparing the leading order behavior of the relevant terms,
only this shortcut will be followed here.

Because of the additional factors due to matter, the low-energy
behavior of the system can be quite different than before.
Compare, for instance, the partition functions (\ref{ZHfattorus})
and (\ref{ZHthintorus}) associated with homogeneous graphs. The
contribution due to the isotropic torus can be much larger than
the contribution from long-and-thin torus if (recall that
$a\epsilon_H < \epsilon_H < 0$) \be \epsilon_H < a\epsilon_H +
\frac{1}{8m\ell^2}+\frac{1}{2\beta}\ln \frac{8m\ell^2}{\beta}. \ee
The last term on the right hand side becomes negligible if
$\ell^2$ is small and $\beta$ large (as it is expected to be if it
scales logarithmically with $N$). The remaining inequality may be
satisfied by judiciously choosing the parameters $m$, $\ell$, and
the coupling $g_B$ appearing inside the energy $\epsilon_H$. Thus,
matter can select the isotropic graph in the low-temperature
regime even though the long-and-thin configuration contains more
cycles. Next, comparing quantities (\ref{ZHdtorus}) for various
$d$, one can find the dimension $d$ of the most optimal torus.
This dimension $d$ will be primarily determined by the number of
short cycles, i.e. by the numbers $a_d$.

Suppose that the optimal dimension is $d=2$ so that the dominant
homogeneous term is given by (\ref{ZHfattorus}). Compare this to
the contribution (\ref{ZLowmatter}) from the low-energy chainlike
graphs. The homogeneous graph can be dominant if the couplings are
chosen such that (recall that in general $\epsilon_0 < \epsilon_H
< 0$) \be \epsilon_H < \epsilon_0 +
\frac{1}{8m_L\ell^2}+\frac{1}{\beta}\ln \frac{8m\ell^2}{\beta}.
\ee If this is satisfied, then low-energy behavior of the graph
system with matter is drastically different from what was found in
Sec. \ref{s_statmech1}. Also, as the matter contribution to the
partition function becomes important, the cycle counting becomes
less so. This implies that the underlying graph structure, while
preserving the large-scale properties of the emergent geometry,
can perhaps become more fluid at small scales than the figures of
homogeneous lattices might imply.

In the high-energy regime, just as before, random graphs can
dominate above some transition temperature. The transition
temperature to this regime should again be proportional to $\ln
N$. The estimate of the coefficient in front of this factor should
depend on the effective masses $m$ and $m_R$ of the particles in
the high and low-temperature regimes. Since the matter factor
associated with random graphs are not known in detail, this
transition is not discussed further here.


\section{Discussion \label{s_discussion}}

Summarizing, this paper focused on the classical statistical
mechanics of graphity models and produced two types of results.
First, results from Monte Carlo simulations show that neither
$r>0$ or $r<0$ basic models produce ground state graphs that are
latticelike. The more interesting case with negative $r$ can at
best produce an extended graph that is made up of smaller units
chained together in a circle. Thus it appears that the graphity
Hamiltonian that depends on vertex degrees and counts closed
cycles must be supplemented with some additional conditions.
Indeed, ground states searched for when the original graphity
model was introduced were implicitly assumed to be homogeneous or
almost homogeneous \cite{graphity1,graphity2}. Simulations of
extended models implementing such homogeneity conditions in turn
reveal that the intuitions in the early works
\cite{graphity1,graphity2} are correct and that emergent graphs in
the high inverse temperature regime can really describe extended
geometries.

Second, the partition functions for the models with and without
matter are estimated analytically to better understand the
numerical results. In the context of the basic model without
matter, such estimates, while very rough, can nonetheless provide
useful information about the behavior of the graphity models in
the continuum $N\ra \infty$ limit. If this limit is defined
through a family of models with constant couplings, the transition
inverse temperature $\beta$ can be shown to scale logarithmically
with $N$. Therefore, the transition occurs at zero temperature for
large $N$. (The transition temperature could be finite if, for
example, the couplings were taken to run logarithmically with $N$
instead.) In the context of the model including matter, estimates
of the extended partition function explain how matter degrees of
freedom can alter the low and high-energy behavior of a graphity
model. In particular, they show that matter can enhance the
importance of homogeneous graph configurations in the partition
function to the extent that these graphs become dominant.
Unfortunately, it was not possible to test the particular form of
matter used in these estimates in numerical simulations.

The problem with nonextended configurations in the large inverse
temperature regime in the graphity models is reminiscent of
similar findings in the context of Euclidean dynamical
triangulations \cite{DynTrianLorentzQG}. In that context, the
problem was noticed to become milder when the triangulation was
coupled to gauge fields \cite{Bilke1,Bilke2}, but the main
solution has been based on restricting the configuration space to
those triangulations respecting certain criteria motivated by
causality \cite{DynTrianLorentzQG} and the presence of global
spatial hypersurfaces \cite{Konopka:2005ps}. In graphity models,
such a restriction of the configurations space cannot be naturally
introduced. Instead, the proposal to bypass the problem of
nonextended geometries is to make the configuration space larger
to include matter degrees of freedom. This is close in spirit to
the mentioned work on dynamical triangulations with matter fields
\cite{Bilke1,Bilke2}; the very interesting question of whether the
mechanism described here might also regularize Euclidean dynamical
triangulations deserves some further thought.

In the end, what can be learned from simple models based on
graphs? At the very least, they should be seen as concrete
examples of setups that do not require a continuous manifold as an
ingredient but rather are able to produce one (or an approximation
to one) in an appropriate regime. In particular, they can produce
flat emergent geometries in the form of tori. Given this, one can
start asking if some graphity model can actually reproduce the
known gravitational dynamics in three plus one dimensions in some
region of its parameter space, or if the discreteness may have
observable consequences for particle physics or cosmology (works
studying related issues in similar models include
\cite{Dowker:2003hb,Magueijo:2007wf}.) These are issues that are
worth exploring further and now bring graphity models, which begin
at a slightly speculative starting point, possibly close to being
compared with experiment.

\bigskip

{\bf Acknowledgments. } I would like to thank my colleagues for
discussions and comments on the manuscript. I am very grateful to
G. Barkema, in particular, for practical programming tips leading
to speeding up cycle counting and for advice regarding Monte Carlo
simulations in general.

\end{document}